\documentclass[draft]{agujournal2019}
\usepackage{lmodern}
\usepackage{url}
\usepackage[inline]{trackchanges}
\usepackage{soul}
\usepackage{natbib}
\usepackage{graphicx}
\usepackage{lineno}
\usepackage{booktabs}
\usepackage{siunitx}
\usepackage{wrapfig}
\draftfalse
\journalname{Geophysical Research Letters}
\begin{document}
\nolinenumbers
\title{A First Look at Hydrogen Generation in an Ultramafic Rock with Micro-CT and SEM-BEX}
\authors{Menke, Hannah P.\affil{1,3}*; Jangda, Zaid Z.\affil{2,3}; Jones, CJ\affil{1,3}; Webb, Max\affil{1,3}; Buckman, Jim\affil{1}; Gough, Amy\affil{1,3}}
\affiliation{1}{Institute of GeoEnergy Engineering, Heriot-Watt University, Edinburgh, UK}
\affiliation{2}{The Lyell Centre, Heriot-Watt University, Edinburgh, UK}
\affiliation{3}{Subsurface Energy Transition \& Innovation Centre, Heriot-Watt University, Edinburgh, UK}
\correspondingauthor{*Hannah P. Menke}{h.menke@hw.ac.uk}
\begin{keypoints}
\item 4D X-ray micro-CT captures in situ gas exsolution during reaction of ultramafic grains.
\item Gas emergence is delayed, followed by rapid bubble growth and evolving gas connectivity.
\item Independent hydrogen measurements and a companion reactor experiment confirm H$_2$ generation.
\end{keypoints}

\begin{abstract}
We present a 4D (time-lapse) micro-CT experiment that visualizes gas emergence and growth during water--rock reaction in an ultramafic granular pack at 100$^\circ$C. Repeated scans resolve the onset of gas exsolution after an initial delay, followed by rapid bubble growth, coalescence, and evolving gas connectivity. Image-derived metrics quantify changes in bulk gas saturation and gas clustering through time. Hydrogen was detected independently in the pump effluent and by an indicator strip in the confining sleeve. A companion reactor experiment using the same parent rock at 100$^\circ$C yielded an estimated net production of 0.65 mmol H$_2$ over approximately three weeks, accounting for dissolved hydrogen and gas removed during sampling. Complementary SEM and multi-element backscattered electron and X-ray (BEX) imaging of pre- and post-reaction grains indicates localized alteration textures consistent with reaction-driven H$_2$ production. These observations constrain pore-scale gas accumulation and multiphase transport relevant to natural hydrogen accumulations.
\end{abstract}

\section*{Plain Language Summary}
Hydrogen gas can form naturally when water reacts with iron-bearing rocks, but measurements of collected gas do not show how bubbles develop inside the rock. We ran a proof-of-concept laboratory experiment in which crushed ultramafic rock was reacted with degassed salty water at 100$^\circ$C while we repeatedly scanned the sample using X-ray micro-computed tomography. The scans show when gas bubbles first appear, how fast they grow, and whether they connect into larger clusters. Hydrogen was detected after the imaging experiment, and a separate reactor experiment using the same rock yielded an estimated net production of approximately 15 mL of hydrogen at standard temperature and pressure over three weeks. We then examined grains before and after the imaging experiment using electron microscopy to investigate changes at their surfaces. Together, these measurements show how hydrogen generation can accompany gas accumulation in porous materials, helping to inform models of where natural hydrogen might be generated, trapped, and detected in the subsurface.

\section{Introduction}
Natural geologic hydrogen (H$_2$) is attracting interest as a potential primary energy resource. Global compilations of gas analyses from seeps, boreholes and mines show that H$_2$ occurs in many tectonic settings, often as a major component of the gas phase \citep{Zgonnik2020,Wong2026}. Recent system-scale syntheses formalise the concept of an H$_2$ system, in which H$_2$ is generated, migrates and accumulates in traps \citep{Jackson2024}, and reviews examine migration pathways and timescales from the planetary to basin scale \citep{Lodhia2024_H2Migration}. This has led to growing interest in natural hydrogen associated with alteration-driven systems \citep{Gaucher2020}, and to proposals for engineered rock-based hydrogen generation that exploit water--rock reactions in mafic and ultramafic rocks \citep{Osselin2022}.

Experimental studies constrain the rates and controls of H$_2$ generation, commonly using batch or flow-through reactors containing powders or finely crushed rock. Low-temperature experiments on partially serpentinized peridotite and olivine show H$_2$ production over days to months \citep{Mayhew2013,Miller2016,Neubeck2014}. Laboratory studies have examined the influence of grain size, temperature, mineralogy and fluid chemistry on reaction rates and hydrogen yields \citep{Mccollom2016,Ellison2021,Emera2026,Wu2026}. Flow-through experiments on olivine have demonstrated sustained H$_2$ generation during fluid circulation \citep{Ross2025_OlivineFlow}, and recent experiments have reported hydrogen generation from stimulated, fractured ultramafic rock \citep{GalvisSilva2025,Shabdirova2027}. These studies underpin parameterisations of H$_2$ generation and identify controls associated with reactive Fe-bearing phases and secondary mineral formation \citep{Schrenk2013}. Bulk measurements, however, do not reveal where gas first emerges within the pore network or how pore geometry controls the transition from dissolved to free gas.

Core-scale reactive transport experiments preserve rock fabric and permeability while allowing geochemical monitoring. Percolation of reactive fluids through ultramafic rocks and serpentinites has demonstrated coupling between alteration and permeability \citep{Wang2019,Albers2021,Farough2016}. In most such studies, gas production is inferred from outlet fluid chemistry or headspace analyses, while the distribution of gas inside the pore network remains unobserved. The pore-scale mechanisms governing the onset of gas emergence and the development of gas connectivity therefore remain poorly constrained.

In parallel, X-ray micro-tomography has been used to image dynamic multiphase flow and reactive transport in rocks. Time-resolved micro-CT has been applied to dissolution and precipitation in carbonates and sandstones, enabling direct observations of evolving pore structures and reaction fronts \citep{NoirielSoulaine2021,NoirielRenard2022,Menke2015}. Similar techniques have quantified residual trapping and ganglion dynamics during CO$_2$--brine flow and constrained pore-scale transport models \citep{Andrew2014,Jiang2015,Bultreys2016}. Recent experiments have also combined hydrogen measurements with four-dimensional X-ray CT observations of mineral replacement and serpentine vein growth in peridotite \citep{Xia2026}.

Micro-CT studies of externally supplied H$_2$ resolve gas displacement and trapping within reservoir rocks \citep{Thaysen2023,Jangda2023,Dokhon2026}. Complementary core-flooding and relative-permeability measurements quantify H$_2$--water flow and its dependence on gas saturation \citep{Yekta2018,Rezaei2022,Boon2022}. These studies provide benchmarks for underground H$_2$ storage, where dissolution, pore-scale trapping and reactions with the rock and microbiota influence containment and recovery \citep{Heinemann2021,Miocic2023}. Reaction-generated gas raises a different question: how does an initially liquid-saturated pore network evolve as hydrogen accumulates and a separate gas phase develops?

In this study, we combine an \emph{in situ} 4D micro-CT experiment with pre- and post-reaction scanning electron microscopy and backscattered electron and X-ray (SEM-BEX) analyses of an ultramafic granular pack reacting with brine at 100$^\circ$C and 4 bar pore pressure. Our approach allows us to document the transition from a liquid-saturated pore network to the appearance and growth of a gas phase, characterise changes in gas saturation and connectivity, and relate these observations to textural and compositional differences at grain surfaces. Independent hydrogen measurements in the imaging apparatus and a companion reactor experiment using the same parent rock establish hydrogen generation alongside the pore-scale observations. Together, these experiments link water--rock reaction with the emergence and accumulation of gas in ultramafic pore space.

\section{Materials and Methods}
We performed a proof-of-concept, time-lapse 4D micro-CT experiment to observe gas exsolution during water--rock reaction in an ultramafic granular pack at 100$^\circ$C. Sample PS23-011, a serpentinized lherzolite with relict olivine from the Cyclops Ophiolite, Indonesia (Supporting Information Text S1 and Figure S1), was crushed and sieved to 0.6--1 mm, packed in an X-ray-transparent Hassler-type holder and saturated with degassed 5 wt.\% KI brine following an initial CO$_2$ flush and extensive brine flooding. The system was held without imposed through-flow at 4 bar pore pressure and 10 bar confining pressure during repeated scans over approximately 30 h. Pressure control accommodated fluid displacement. Each acquisition lasted approximately 50 min and had a voxel size of 6.277 $\mu$m. Gas, brine-filled pore space and solid were segmented from the registered volumes to quantify the onset of gas emergence, resolved bulk gas saturation and the fraction of gas in the largest connected cluster. After the experiment, H$_2$ was measured in the pump gas effluent during depressurisation using a portable electrochemical detector, and the H$_2$-indicator strip in the confining sleeve was inspected.

The companion reactor experiment used approximately 200 g of crushed PS23-011 and 430 mL of degassed deionised water in a 600 mL vessel. The rock displaced approximately 70 mL, leaving a nominal 100 mL helium headspace. The reactor was maintained at 100$^\circ$C and nominally 4 bar gauge for approximately three weeks, with helium supplied by a pump to maintain pressure during gas sampling. Headspace gas was analysed by mass spectrometry at $m/z=2$, 4, 18, 28, 32 and 44, and gas mole fractions were estimated using instrument-specific relative sensitivity factors for H$_2$, He and CO$_2$, with literature values retained for H$_2$O and O$_2$ and N$_2$ as the reference. The final sample, collected after depressurisation, was excluded from the pressurised time series.

Pre- and post-experiment grains from the imaging study were examined using SEM and multi-element BEX imaging. Pre- and post-experiment observations were made on separate grain mounts, with five elemental spot analyses per mount. Full protocols, the supplied reactor signals, processing assumptions, sampling records and additional image metrics are provided in the Supporting Information.

\section{Results}
\subsection{Gas emergence and pore occupancy}
The 4D micro-CT scans provided a direct visualization of gas-phase evolution within the pore space. During the initial heating and early static reaction period, no gas was resolved in the brine-filled pore space. The image-derived porosity was 0.375 and the specific solid--pore interfacial area was $1.13\times10^4\ \mathrm{m^2\,m^{-3}}$ (Table S1a).

\begin{figure}
\centering
\includegraphics[width=\linewidth]{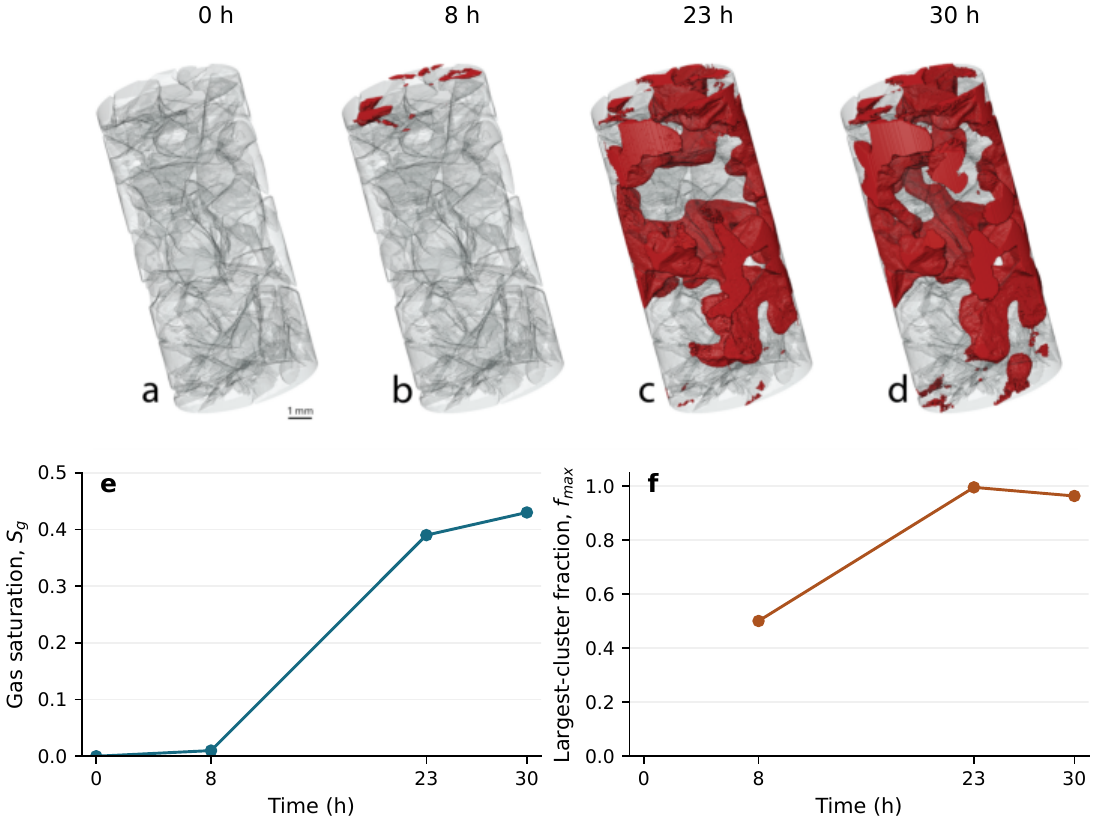}
\caption{Time-resolved gas evolution in the ultramafic granular pack. (a--d) Segmented gas (red) at 0, 8, 23 and 30 h, with translucent grains and transparent brine. Scale bar: 1 mm. (e) Gas saturation of the imaged volume. (f) Fraction of segmented gas in the largest connected cluster, shown for scans containing gas.}
\label{fig:ImageResults}
\end{figure}

Gas was first resolved at 8 h, when discrete bubbles appeared primarily near the top of the imaged interval (Figure~\ref{fig:ImageResults}). The bulk gas saturation remained low, at approximately 0.01. The delayed appearance of a gas phase is consistent with an initial period of dissolved hydrogen accumulation before gas exsolution.

By 23 h, gas extended throughout much of the imaged volume and $S_g$ had increased to approximately 0.39. At 30 h, $S_g$ reached approximately 0.43. The reported interval-average changes in saturation were approximately $2.56\times10^{-2}$ h$^{-1}$ between 8 and 23 h and $5.87\times10^{-3}$ h$^{-1}$ between 23 and 30 h, a reduction by a factor of approximately 4.4. These rates quantify the increase in gas saturation within the imaged pore volume.

The largest-cluster fraction increased from approximately 0.501 at 8 h to 0.996 at 23 h and was 0.964 at 30 h. Thus, most segmented gas belonged to one large domain at the later times, although some redistribution among clusters remained apparent. Vertical saturation profiles show a heterogeneous gas distribution along the imaged column (Figure S4). At 8 h, resolved gas was concentrated near the top. By 23 h, gas occupied much of the column, with the highest local saturation around 4 mm depth. The 30 h profile retained substantial gas at depth while local saturations changed. Gas at depth is consistent with in situ formation and capillary retention within the granular pore network.

During controlled depressurisation after imaging, the electrochemical detector registered H$_2$ concentrations up to approximately 80 ppm in the pump gas effluent. The receiving pump also contained CO$_2$ retained from the initial core flush, so the measured H$_2$ concentration applies to diluted effluent gas rather than undiluted pore gas. The H$_2$-indicator strip in the confining sleeve, inspected after the experiment, also showed a positive response. These independent observations support hydrogen generation during the imaging experiment.

\subsection{Hydrogen generation in the companion reactor}
\begin{figure}
\centering
\includegraphics[width=\linewidth]{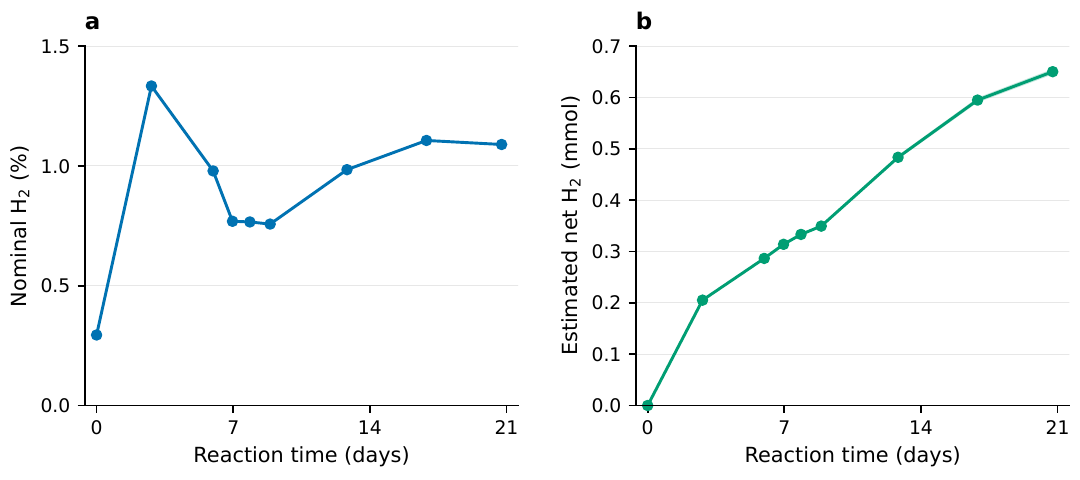}
\caption{Hydrogen generation in the companion reactor at 100$^\circ$C and nominally 4 bar gauge. (a) Nominal H$_2$ mole fraction calculated from the supplied mass spectra using the relative sensitivity factors described in Supporting Information Text S2. (b) Estimated cumulative net H$_2$ production since the first sample, including dissolved hydrogen and sampling losses. The shaded interval spans calculations for sampling-loop temperatures of 20 and 25$^\circ$C.}
\label{fig:reactor}
\end{figure}

The companion reactor experiment provided additional evidence of hydrogen generation using material from the same parent rock. Estimated net hydrogen production over approximately three weeks was 0.65 mmol H$_2$, equivalent to approximately 15 mL at STP (0$^\circ$C and 1 atm), or 3.3 $\mu$mol per gram of rock (Figure~\ref{fig:reactor}). The mass balance includes hydrogen in the headspace and dissolved in the water, accounts for hydrogen removed during sampling, and subtracts the initial inventory. The calculation is detailed in the Supporting Information.

Nominal sampled-gas H$_2$ mole fractions were 0.29\%, 1.33\% and 1.09\% in the first, second and final pressurised samples, respectively (Figure~\ref{fig:reactor}; Table S4).

\subsection{Grain-surface observations}
\begin{figure}
\centering
\includegraphics[width=\linewidth]{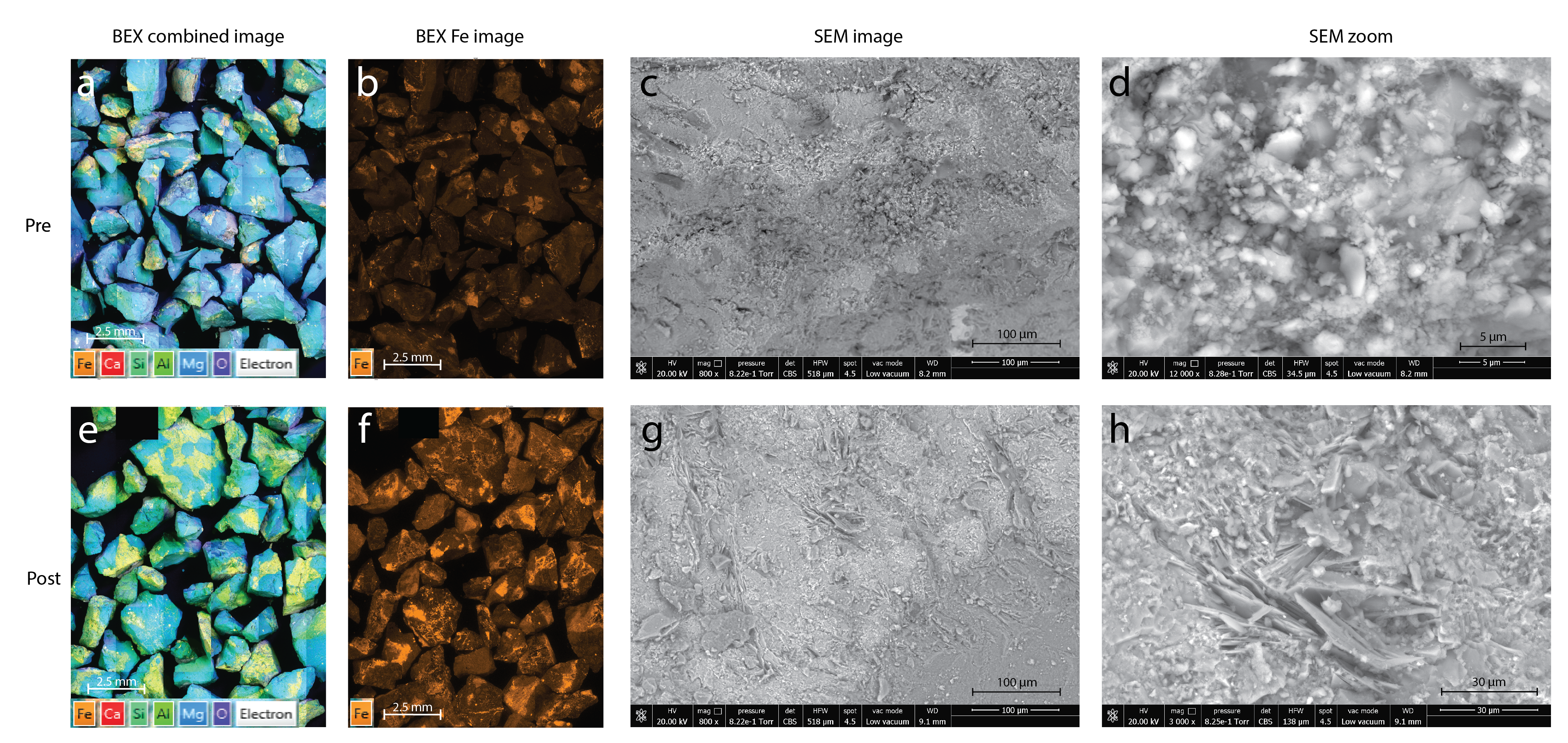}
\caption{Pre- and post-experiment grain mounts from the imaging study. (a,e) Multi-element BEX composites. (b,f) Fe maps. (c,g) SEM overviews. (d,h) Higher-magnification views showing compact surface material before the experiment and elongate or platy features on a post-experiment grain. Pre- and post-experiment panels show different grains, and (d) and (h) are at different magnifications.}
\label{fig:SEM}
\end{figure}

BEX composite images show that both pre- and post-reaction grain mounts are compositionally heterogeneous (Figure~\ref{fig:SEM}a,e). In the post-experiment sample, patches of higher intensity occur on a subset of grains in both the multi-element composite and the Fe maps (Figure~\ref{fig:SEM}b,f).

Pre-experiment grain surfaces are relatively smooth at the overview scale and consist mainly of compact, equant to sub-rounded microcrystalline material at higher magnification (Figure~\ref{fig:SEM}c,d). In contrast, post-experiment grains show more irregular, roughened surfaces and discrete clusters of elongate to platy features (Figure~\ref{fig:SEM}g,h). These textures are consistent with local surface alteration or exposure of previously obscured domains. Different grains were examined before and after reaction, and drying without a KI-removal rinse may also have contributed surface residues.

Five spot analyses on each mount indicated overlapping Mg- and Fe-bearing compositions and some post-experiment spots with higher Al and Ca (Table S2). Post-experiment Mg values extended to lower concentrations than those measured before the experiment. One possible explanation is local dissolution of Mg-rich material, leaving or exposing relatively Al- and Ca-enriched domains. Together with the patchy surface textures, these compositional differences are consistent with heterogeneous alteration at grain surfaces.

\section{Discussion}
\label{sec:Discussion}
The pore-scale images show three stages of gas evolution: an initial period with no resolvable gas, a rapid increase in gas saturation between 8 and 23 h, and more modest growth between 23 and 30 h. Simple mass-balance and transport estimates help to interpret this sequence in terms of dissolved hydrogen accumulation, gas exsolution and subsequent bubble growth.

The delay before the first visible bubbles can be understood as the time needed to build up dissolved H$_2$ in the pore water. A mass balance based on the first gas-bearing scan estimates that the brine reached hydrogen saturation after approximately six hours, assuming constant hydrogen generation during the first eight hours and equilibrium partitioning between brine and gas. The derivation accounts for dissolved and free hydrogen and for hydrogen exported in brine displaced by gas expansion (Supporting Information). Gas domains below the imaging resolution may precede the first resolvable bubbles. At 100$^\circ$C, the 4 bar pore pressure suppresses bulk boiling, although water vapour can contribute to the gas phase.

The rapid increase in gas saturation from approximately 0.01 to 0.39 between 8 and 23 h is consistent with a period of widespread gas emergence and growth. During this interval, newly generated H$_2$ can partition between solution and existing bubbles. Assuming a reference diffusion coefficient for dissolved H$_2$ in water of $D=10^{-9}\ \mathrm{m^2\,s^{-1}}$, the diffusive length scale is
\begin{equation}
\ell_D\sim\sqrt{Dt}\approx5\mbox{--}9\ \mathrm{mm}
\end{equation}
for $t$ between 8 and 23 h. Dissolved hydrogen can therefore be redistributed over millimetre distances and feed bubble growth between scans. By 23 h, approximately 99.6\% of the segmented gas belonged to a single connected cluster.

Between 23 and 30 h, gas saturation increased from approximately 0.39 to 0.43, while its interval-average rate of increase fell by a factor of approximately 4.4. This slower increase may reflect changes in hydrogen generation, gas--liquid partitioning or transport as the gas phase develops. The post-experiment surface textures are also consistent with alteration that could change local reactivity. Gas saturation records the gas retained within the imaged pore space, so continued generation need not produce a proportional increase in local gas volume.

Independent hydrogen measurements connect these pore-scale observations with water--rock reaction. The electrochemical detector and indicator strip detected hydrogen in the imaging apparatus, and the companion reactor generated an estimated net 0.65 mmol H$_2$. The reactor contained a helium headspace into which hydrogen could partition, whereas the imaging experiment began with brine-saturated pores. Both experiments used the same nominal temperature and pressure, but differed in fluid composition, initial gas space, duration and sampling. They therefore examine hydrogen generation under related conditions and its accumulation in different gas--liquid configurations.

The results show that a substantial lag can occur between the onset of water--rock reaction and the appearance of a resolvable gas phase, followed by a shorter interval of rapid gas accumulation and increasing connectivity. In natural H$_2$ systems, hydrogen may accumulate in solution before a free-gas phase becomes detectable. Local pore geometry and capillary retention will influence whether that gas remains trapped or contributes to migration. The time-resolved images provide the spatial observations needed to examine these transitions alongside chemical measurements of hydrogen generation.

\section{Conclusions}
This proof-of-concept experiment demonstrates that time-resolved 4D micro-CT can monitor the emergence and early evolution of a gas phase during water--rock reaction in an ultramafic granular pack at 100$^\circ$C. After an initial period of approximately eight hours with no resolvable gas, gas saturation increased rapidly to approximately 0.39 at 23 h and then more slowly to 0.43 at 30 h. Most late-time gas belonged to one connected cluster. A mass-balance reconstruction gives a saturation timescale of approximately six hours, consistent with dissolved hydrogen accumulation before gas emergence.

Hydrogen generation was established by independent measurements in the imaging apparatus and mass spectrometry in a companion reactor containing the same parent rock. The reactor generated an estimated net 0.65 mmol H$_2$ over approximately three weeks, accounting for hydrogen dissolved in the water and removed during sampling. SEM-BEX observations are consistent with local surface alteration alongside the development of the gas phase.

Taken together, the results show how hydrogen generation can precede the appearance of a resolvable gas phase and how gas connectivity can develop over relatively short timescales. This behaviour has implications for natural H$_2$ systems and underground H$_2$ storage, where partitioning between dissolved, trapped and mobile gas controls detection, migration and recoverability. Future work should extend this approach across a broader range of mineralogies, pressures, temperatures and flow regimes to quantify the coupling between water--rock reaction, gas generation and multiphase transport.

\section*{Supporting Information}
Supporting Information provides expanded experimental methods, hydrogen mass-balance calculations, supplementary imaging and rock composition data, and reactor mass spectrometry signals and sampling records.

\section*{Open Research Section}
The micro-CT image volumes, processed segmentations, and representative SEM-BEX images used in this study are available in an open-access repository on Zenodo \citep{Menke2025Dataset}. The baseline-corrected reactor mass spectrometry signals, relative sensitivity factors and sampling-event records are provided in the Supporting Information.

\section*{Conflict of Interest declaration}
The authors declare there are no conflicts of interest for this manuscript.

\acknowledgments
This research was funded by The Carnegie Trust RIG013387. Sample collection was funded by the Gilchrist Educational Trust (grant GILC 06.22, administered by the Royal Geographical Society).

\section*{Author Contributions}
H.P.M and A.G. conceptualised this study. H.P.M. raised funding. A.G. chose the sample. M.W. prepared the sample. H.P.M. and Z.Z.J. conducted the imaging experiment. J.B. conducted the pre- and post-experiment SEM-BEX imaging. Z.Z.J. and H.P.M. analysed the micro-CT image data. M.W. analysed the SEM-BEX data. C.J.J. conducted the companion reactor experiment and processed the mass spectrometer data. H.P.M. wrote the first draft. All authors contributed to writing this paper.
\nocite{Buades2005,Lanczos1956,IAPWS2004Henry,IAPWS1992Saturation}
\bibliography{references}

\begin{thebibliography}{}

\bibitem [\protect \citeauthoryear {%
Albers%
, Bach%
, P{\'e}rez-Gussiny{\'e}%
, McCammon%
\BCBL {}\ \BBA {} Frederichs%
}{%
Albers%
\ \protect \BOthers {.}}{%
{\protect \APACyear {2021}}%
}]{%
Albers2021}
\APACinsertmetastar {%
Albers2021}%
\begin{APACrefauthors}%
Albers, E.%
, Bach, W.%
, P{\'e}rez-Gussiny{\'e}, M.%
, McCammon, C.%
\BCBL {}\ \BBA {} Frederichs, T.%
\end{APACrefauthors}%
\unskip\
\newblock
\APACrefYearMonthDay{2021}{}{}.
\newblock
{\BBOQ}\APACrefatitle {Serpentinization-driven {H$_2$} production from
  continental break-up to mid-ocean ridge spreading: unexpected high rates at
  the {West Iberia} margin} {Serpentinization-driven {H$_2$} production from
  continental break-up to mid-ocean ridge spreading: unexpected high rates at
  the {West Iberia} margin}.{\BBCQ}
\newblock
\APACjournalVolNumPages{Frontiers in Earth Science}{9}{}{673063}.
\newblock
\begin{APACrefDOI} \doi{10.3389/feart.2021.673063} \end{APACrefDOI}
\PrintBackRefs{\CurrentBib}

\bibitem [\protect \citeauthoryear {%
Andrew%
, Bijeljic%
\BCBL {}\ \BBA {} Blunt%
}{%
Andrew%
\ \protect \BOthers {.}}{%
{\protect \APACyear {2014}}%
}]{%
Andrew2014}
\APACinsertmetastar {%
Andrew2014}%
\begin{APACrefauthors}%
Andrew, M.%
, Bijeljic, B.%
\BCBL {}\ \BBA {} Blunt, M\BPBI J.%
\end{APACrefauthors}%
\unskip\
\newblock
\APACrefYearMonthDay{2014}{}{}.
\newblock
{\BBOQ}\APACrefatitle {Pore-scale imaging of trapped supercritical carbon
  dioxide in sandstones and carbonates} {Pore-scale imaging of trapped
  supercritical carbon dioxide in sandstones and carbonates}.{\BBCQ}
\newblock
\APACjournalVolNumPages{International Journal of Greenhouse Gas
  Control}{22}{}{1--14}.
\newblock
\begin{APACrefDOI} \doi{10.1016/j.ijggc.2013.12.018} \end{APACrefDOI}
\PrintBackRefs{\CurrentBib}

\bibitem [\protect \citeauthoryear {%
Boon%
\ \BBA {} Hajibeygi%
}{%
Boon%
\ \BBA {} Hajibeygi%
}{%
{\protect \APACyear {2022}}%
}]{%
Boon2022}
\APACinsertmetastar {%
Boon2022}%
\begin{APACrefauthors}%
Boon, M.%
\BCBT {}\ \BBA {} Hajibeygi, H.%
\end{APACrefauthors}%
\unskip\
\newblock
\APACrefYearMonthDay{2022}{}{}.
\newblock
{\BBOQ}\APACrefatitle {Experimental characterization of {H$_2$}/water
  multiphase flow in heterogeneous sandstone rock at the core scale relevant
  for underground hydrogen storage ({UHS})} {Experimental characterization of
  {H$_2$}/water multiphase flow in heterogeneous sandstone rock at the core
  scale relevant for underground hydrogen storage ({UHS})}.{\BBCQ}
\newblock
\APACjournalVolNumPages{Scientific Reports}{12}{1}{14604}.
\newblock
\begin{APACrefDOI} \doi{10.1038/s41598-022-18759-8} \end{APACrefDOI}
\PrintBackRefs{\CurrentBib}

\bibitem [\protect \citeauthoryear {%
Buades%
, Coll%
\BCBL {}\ \BBA {} Morel%
}{%
Buades%
\ \protect \BOthers {.}}{%
{\protect \APACyear {2005}}%
}]{%
Buades2005}
\APACinsertmetastar {%
Buades2005}%
\begin{APACrefauthors}%
Buades, A.%
, Coll, B.%
\BCBL {}\ \BBA {} Morel, J\BHBI M.%
\end{APACrefauthors}%
\unskip\
\newblock
\APACrefYearMonthDay{2005}{}{}.
\newblock
{\BBOQ}\APACrefatitle {A non-local algorithm for image denoising} {A non-local
  algorithm for image denoising}.{\BBCQ}
\newblock
\BIn{} \APACrefbtitle {2005 {IEEE Computer Society} Conference on Computer
  Vision and Pattern Recognition ({CVPR}'05)} {2005 {IEEE Computer Society}
  conference on computer vision and pattern recognition ({CVPR}'05)}\ (\BVOL~2,
  \BPGS\ 60--65).
\newblock
\begin{APACrefDOI} \doi{10.1109/cvpr.2005.38} \end{APACrefDOI}
\PrintBackRefs{\CurrentBib}

\bibitem [\protect \citeauthoryear {%
Bultreys%
, De~Boever%
\BCBL {}\ \BBA {} Cnudde%
}{%
Bultreys%
\ \protect \BOthers {.}}{%
{\protect \APACyear {2016}}%
}]{%
Bultreys2016}
\APACinsertmetastar {%
Bultreys2016}%
\begin{APACrefauthors}%
Bultreys, T.%
, De~Boever, W.%
\BCBL {}\ \BBA {} Cnudde, V.%
\end{APACrefauthors}%
\unskip\
\newblock
\APACrefYearMonthDay{2016}{}{}.
\newblock
{\BBOQ}\APACrefatitle {Imaging and image-based fluid transport modeling at the
  pore scale in geological materials: A practical introduction to the current
  state-of-the-art} {Imaging and image-based fluid transport modeling at the
  pore scale in geological materials: A practical introduction to the current
  state-of-the-art}.{\BBCQ}
\newblock
\APACjournalVolNumPages{Earth-Science Reviews}{155}{}{93--128}.
\newblock
\begin{APACrefDOI} \doi{10.1016/j.earscirev.2016.02.001} \end{APACrefDOI}
\PrintBackRefs{\CurrentBib}

\bibitem [\protect \citeauthoryear {%
Dokhon%
, Bijeljic%
\BCBL {}\ \BBA {} Blunt%
}{%
Dokhon%
\ \protect \BOthers {.}}{%
{\protect \APACyear {2026}}%
}]{%
Dokhon2026}
\APACinsertmetastar {%
Dokhon2026}%
\begin{APACrefauthors}%
Dokhon, W.%
, Bijeljic, B.%
\BCBL {}\ \BBA {} Blunt, M\BPBI J.%
\end{APACrefauthors}%
\unskip\
\newblock
\APACrefYearMonthDay{2026}{}{}.
\newblock
{\BBOQ}\APACrefatitle {The combined effects of pressure decline and gas
  withdrawal in underground hydrogen storage: A pore-scale experimental study}
  {The combined effects of pressure decline and gas withdrawal in underground
  hydrogen storage: A pore-scale experimental study}.{\BBCQ}
\newblock
\APACjournalVolNumPages{Advances in Geo-Energy Research}{20}{1}{71--84}.
\newblock
\begin{APACrefDOI} \doi{10.46690/ager.2026.04.06} \end{APACrefDOI}
\PrintBackRefs{\CurrentBib}

\bibitem [\protect \citeauthoryear {%
Ellison%
\ \protect \BOthers {.}}{%
Ellison%
\ \protect \BOthers {.}}{%
{\protect \APACyear {2021}}%
}]{%
Ellison2021}
\APACinsertmetastar {%
Ellison2021}%
\begin{APACrefauthors}%
Ellison, E\BPBI T.%
, Templeton, A\BPBI S.%
, Zeigler, S\BPBI D.%
, Mayhew, L\BPBI E.%
, Kelemen, P\BPBI B.%
, Matter, J\BPBI M.%
\BCBL {}\ \BBA {} {The Oman Drilling Project Science Party}.%
\end{APACrefauthors}%
\unskip\
\newblock
\APACrefYearMonthDay{2021}{}{}.
\newblock
{\BBOQ}\APACrefatitle {Low-temperature hydrogen formation during aqueous
  alteration of serpentinized peridotite in the {Samail} ophiolite}
  {Low-temperature hydrogen formation during aqueous alteration of
  serpentinized peridotite in the {Samail} ophiolite}.{\BBCQ}
\newblock
\APACjournalVolNumPages{Journal of Geophysical Research: Solid
  Earth}{126}{6}{e2021JB021981}.
\newblock
\begin{APACrefDOI} \doi{10.1029/2021jb021981} \end{APACrefDOI}
\PrintBackRefs{\CurrentBib}

\bibitem [\protect \citeauthoryear {%
Emera%
, Angelina%
, Gyaubea Matey-Korley%
, Isah%
\BCBL {}\ \BBA {} Okoroafor%
}{%
Emera%
\ \protect \BOthers {.}}{%
{\protect \APACyear {2026}}%
}]{%
Emera2026}
\APACinsertmetastar {%
Emera2026}%
\begin{APACrefauthors}%
Emera, R.%
, Angelina, N\BPBI M.%
, Gyaubea Matey-Korley, V.%
, Isah, A.%
\BCBL {}\ \BBA {} Okoroafor, E\BPBI R.%
\end{APACrefauthors}%
\unskip\
\newblock
\APACrefYearMonthDay{2026}{}{}.
\newblock
{\BBOQ}\APACrefatitle {Experiments and geochemical modeling of hydrogen
  generation during water--rock reactions: effects of mineralogy and
  physicochemical conditions} {Experiments and geochemical modeling of hydrogen
  generation during water--rock reactions: effects of mineralogy and
  physicochemical conditions}.{\BBCQ}
\newblock
\APACjournalVolNumPages{Frontiers in Geochemistry}{4}{}{1913188}.
\newblock
\begin{APACrefDOI} \doi{10.3389/fgeoc.2026.1913188} \end{APACrefDOI}
\PrintBackRefs{\CurrentBib}

\bibitem [\protect \citeauthoryear {%
Farough%
, Moore%
, Lockner%
\BCBL {}\ \BBA {} Lowell%
}{%
Farough%
\ \protect \BOthers {.}}{%
{\protect \APACyear {2016}}%
}]{%
Farough2016}
\APACinsertmetastar {%
Farough2016}%
\begin{APACrefauthors}%
Farough, A.%
, Moore, D\BPBI E.%
, Lockner, D\BPBI A.%
\BCBL {}\ \BBA {} Lowell, R\BPBI P.%
\end{APACrefauthors}%
\unskip\
\newblock
\APACrefYearMonthDay{2016}{}{}.
\newblock
{\BBOQ}\APACrefatitle {Evolution of fracture permeability of ultramafic rocks
  undergoing serpentinization at hydrothermal conditions: An experimental
  study} {Evolution of fracture permeability of ultramafic rocks undergoing
  serpentinization at hydrothermal conditions: An experimental study}.{\BBCQ}
\newblock
\APACjournalVolNumPages{Geochemistry, Geophysics, Geosystems}{17}{1}{44--55}.
\newblock
\begin{APACrefDOI} \doi{10.1002/2015gc005973} \end{APACrefDOI}
\PrintBackRefs{\CurrentBib}

\bibitem [\protect \citeauthoryear {%
Galvis~Silva%
, Isah%
, Shabdirova%
, Carrascal%
\BCBL {}\ \BBA {} Okoroafor%
}{%
Galvis~Silva%
\ \protect \BOthers {.}}{%
{\protect \APACyear {2025}}%
}]{%
GalvisSilva2025}
\APACinsertmetastar {%
GalvisSilva2025}%
\begin{APACrefauthors}%
Galvis~Silva, H.%
, Isah, A.%
, Shabdirova, A.%
, Carrascal, F.%
\BCBL {}\ \BBA {} Okoroafor, E\BPBI R.%
\end{APACrefauthors}%
\unskip\
\newblock
\APACrefYearMonthDay{2025}{}{}.
\newblock
{\BBOQ}\APACrefatitle {An Approach to Sustainable Hydrogen Generation From
  Stimulated Iron-Rich Ultramafic Rock} {An approach to sustainable hydrogen
  generation from stimulated iron-rich ultramafic rock}.{\BBCQ}
\newblock
\APACjournalVolNumPages{Geophysical Research Letters}{52}{24}{e2025GL119699}.
\newblock
\begin{APACrefDOI} \doi{10.1029/2025GL119699} \end{APACrefDOI}
\PrintBackRefs{\CurrentBib}

\bibitem [\protect \citeauthoryear {%
Gaucher%
}{%
Gaucher%
}{%
{\protect \APACyear {2020}}%
}]{%
Gaucher2020}
\APACinsertmetastar {%
Gaucher2020}%
\begin{APACrefauthors}%
Gaucher, E\BPBI C.%
\end{APACrefauthors}%
\unskip\
\newblock
\APACrefYearMonthDay{2020}{}{}.
\newblock
{\BBOQ}\APACrefatitle {New perspectives in the industrial exploration for
  native hydrogen} {New perspectives in the industrial exploration for native
  hydrogen}.{\BBCQ}
\newblock
\APACjournalVolNumPages{Elements}{16}{1}{8--9}.
\newblock
\begin{APACrefDOI} \doi{10.2138/gselements.16.1.8} \end{APACrefDOI}
\PrintBackRefs{\CurrentBib}

\bibitem [\protect \citeauthoryear {%
Heinemann%
\ \protect \BOthers {.}}{%
Heinemann%
\ \protect \BOthers {.}}{%
{\protect \APACyear {2021}}%
}]{%
Heinemann2021}
\APACinsertmetastar {%
Heinemann2021}%
\begin{APACrefauthors}%
Heinemann, N.%
, Alcalde, J.%
, Miocic, J\BPBI M.%
, Hangx, S\BPBI J\BPBI T.%
, Kallmeyer, J.%
, Ostertag-Henning, C.%
\BDBL {}Rudloff, A.%
\end{APACrefauthors}%
\unskip\
\newblock
\APACrefYearMonthDay{2021}{}{}.
\newblock
{\BBOQ}\APACrefatitle {Enabling large-scale hydrogen storage in porous
  media--the scientific challenges} {Enabling large-scale hydrogen storage in
  porous media--the scientific challenges}.{\BBCQ}
\newblock
\APACjournalVolNumPages{Energy \& Environmental Science}{14}{2}{853--864}.
\newblock
\begin{APACrefDOI} \doi{10.1039/d0ee03536j} \end{APACrefDOI}
\PrintBackRefs{\CurrentBib}

\bibitem [\protect \citeauthoryear {%
{IAPWS}%
}{%
{IAPWS}%
}{%
{\protect \APACyear {1992}}%
}]{%
IAPWS1992Saturation}
\APACinsertmetastar {%
IAPWS1992Saturation}%
\begin{APACrefauthors}%
{IAPWS}.%
\end{APACrefauthors}%
\unskip\
\newblock
\APACrefYearMonthDay{1992}{{\APACmonth{09}}}{}.
\newblock
\APACrefbtitle {Revised Supplementary Release on Saturation Properties of
  Ordinary Water Substance} {Revised supplementary release on saturation
  properties of ordinary water substance}\ \APACbVolEdTR {}{Supplementary
  Release\ \BNUM\ SR1-86(1992)}.
\newblock
\APACaddressInstitution{}{International Association for the Properties of Water
  and Steam}.
\newblock
\begin{APACrefURL}
  \url{https://iapws.org/technical-guidance/release/Supp-sat.download}
  \end{APACrefURL}
\PrintBackRefs{\CurrentBib}

\bibitem [\protect \citeauthoryear {%
{IAPWS}%
}{%
{IAPWS}%
}{%
{\protect \APACyear {2004}}%
}]{%
IAPWS2004Henry}
\APACinsertmetastar {%
IAPWS2004Henry}%
\begin{APACrefauthors}%
{IAPWS}.%
\end{APACrefauthors}%
\unskip\
\newblock
\APACrefYearMonthDay{2004}{{\APACmonth{09}}}{}.
\newblock
\APACrefbtitle {Guideline on the {Henry's} Constant and Vapor-Liquid
  Distribution Constant for Gases in {H2O} and {D2O} at High Temperatures}
  {Guideline on the {Henry's} constant and vapor-liquid distribution constant
  for gases in {H2O} and {D2O} at high temperatures}\ \APACbVolEdTR
  {}{Guideline\ \BNUM\ G7-04}.
\newblock
\APACaddressInstitution{}{International Association for the Properties of Water
  and Steam}.
\newblock
\begin{APACrefURL} \url{https://iapws.org/documents/release/HenGuide.download}
  \end{APACrefURL}
\PrintBackRefs{\CurrentBib}

\bibitem [\protect \citeauthoryear {%
Jackson%
\ \protect \BOthers {.}}{%
Jackson%
\ \protect \BOthers {.}}{%
{\protect \APACyear {2024}}%
}]{%
Jackson2024}
\APACinsertmetastar {%
Jackson2024}%
\begin{APACrefauthors}%
Jackson, O.%
, Lawrence, S\BPBI R.%
, Hutchinson, I\BPBI P.%
, Stocks, A\BPBI E.%
, Barnicoat, A\BPBI C.%
\BCBL {}\ \BBA {} Powney, M.%
\end{APACrefauthors}%
\unskip\
\newblock
\APACrefYearMonthDay{2024}{}{}.
\newblock
{\BBOQ}\APACrefatitle {Natural hydrogen: sources, systems and exploration
  plays} {Natural hydrogen: sources, systems and exploration plays}.{\BBCQ}
\newblock
\APACjournalVolNumPages{Geoenergy}{2}{1}{geoenergy2024-002}.
\newblock
\begin{APACrefDOI} \doi{10.1144/geoenergy2024-002} \end{APACrefDOI}
\PrintBackRefs{\CurrentBib}

\bibitem [\protect \citeauthoryear {%
Jangda%
\ \protect \BOthers {.}}{%
Jangda%
\ \protect \BOthers {.}}{%
{\protect \APACyear {2023}}%
}]{%
Jangda2023}
\APACinsertmetastar {%
Jangda2023}%
\begin{APACrefauthors}%
Jangda, Z.%
, Menke, H.%
, Busch, A.%
, Geiger, S.%
, Bultreys, T.%
, Lewis, H.%
\BCBL {}\ \BBA {} Singh, K.%
\end{APACrefauthors}%
\unskip\
\newblock
\APACrefYearMonthDay{2023}{}{}.
\newblock
{\BBOQ}\APACrefatitle {Pore-scale visualization of hydrogen storage in a
  sandstone at subsurface pressure and temperature conditions: Trapping,
  dissolution and wettability} {Pore-scale visualization of hydrogen storage in
  a sandstone at subsurface pressure and temperature conditions: Trapping,
  dissolution and wettability}.{\BBCQ}
\newblock
\APACjournalVolNumPages{Journal of Colloid and Interface
  Science}{629}{}{316--325}.
\newblock
\begin{APACrefDOI} \doi{10.1016/j.jcis.2022.09.082} \end{APACrefDOI}
\PrintBackRefs{\CurrentBib}

\bibitem [\protect \citeauthoryear {%
Jiang%
\ \BBA {} Tsuji%
}{%
Jiang%
\ \BBA {} Tsuji%
}{%
{\protect \APACyear {2015}}%
}]{%
Jiang2015}
\APACinsertmetastar {%
Jiang2015}%
\begin{APACrefauthors}%
Jiang, F.%
\BCBT {}\ \BBA {} Tsuji, T.%
\end{APACrefauthors}%
\unskip\
\newblock
\APACrefYearMonthDay{2015}{}{}.
\newblock
{\BBOQ}\APACrefatitle {Impact of interfacial tension on residual {CO$_2$}
  clusters in porous sandstone} {Impact of interfacial tension on residual
  {CO$_2$} clusters in porous sandstone}.{\BBCQ}
\newblock
\APACjournalVolNumPages{Water Resources Research}{51}{3}{1710--1722}.
\newblock
\begin{APACrefDOI} \doi{10.1002/2014wr016070} \end{APACrefDOI}
\PrintBackRefs{\CurrentBib}

\bibitem [\protect \citeauthoryear {%
Lanczos%
}{%
Lanczos%
}{%
{\protect \APACyear {1956}}%
}]{%
Lanczos1956}
\APACinsertmetastar {%
Lanczos1956}%
\begin{APACrefauthors}%
Lanczos, C.%
\end{APACrefauthors}%
\unskip\
\newblock
\APACrefYear{1956}.
\newblock
\APACrefbtitle {Applied Analysis} {Applied analysis}.
\newblock
\APACaddressPublisher{Englewood Cliffs, NJ}{Prentice-Hall}.
\PrintBackRefs{\CurrentBib}

\bibitem [\protect \citeauthoryear {%
Lodhia%
, Peeters%
\BCBL {}\ \BBA {} Frery%
}{%
Lodhia%
\ \protect \BOthers {.}}{%
{\protect \APACyear {2024}}%
}]{%
Lodhia2024_H2Migration}
\APACinsertmetastar {%
Lodhia2024_H2Migration}%
\begin{APACrefauthors}%
Lodhia, B\BPBI H.%
, Peeters, L.%
\BCBL {}\ \BBA {} Frery, E.%
\end{APACrefauthors}%
\unskip\
\newblock
\APACrefYearMonthDay{2024}{}{}.
\newblock
{\BBOQ}\APACrefatitle {A review of the migration of hydrogen from the planetary
  to basin scale} {A review of the migration of hydrogen from the planetary to
  basin scale}.{\BBCQ}
\newblock
\APACjournalVolNumPages{Journal of Geophysical Research: Solid
  Earth}{129}{6}{e2024JB028715}.
\newblock
\begin{APACrefDOI} \doi{10.1029/2024jb028715} \end{APACrefDOI}
\PrintBackRefs{\CurrentBib}

\bibitem [\protect \citeauthoryear {%
Mayhew%
, Ellison%
, McCollom%
, Trainor%
\BCBL {}\ \BBA {} Templeton%
}{%
Mayhew%
\ \protect \BOthers {.}}{%
{\protect \APACyear {2013}}%
}]{%
Mayhew2013}
\APACinsertmetastar {%
Mayhew2013}%
\begin{APACrefauthors}%
Mayhew, L\BPBI E.%
, Ellison, E\BPBI T.%
, McCollom, T\BPBI M.%
, Trainor, T\BPBI P.%
\BCBL {}\ \BBA {} Templeton, A\BPBI S.%
\end{APACrefauthors}%
\unskip\
\newblock
\APACrefYearMonthDay{2013}{}{}.
\newblock
{\BBOQ}\APACrefatitle {Hydrogen generation from low-temperature water--rock
  reactions} {Hydrogen generation from low-temperature water--rock
  reactions}.{\BBCQ}
\newblock
\APACjournalVolNumPages{Nature Geoscience}{6}{6}{478--484}.
\newblock
\begin{APACrefDOI} \doi{10.1038/ngeo1825} \end{APACrefDOI}
\PrintBackRefs{\CurrentBib}

\bibitem [\protect \citeauthoryear {%
McCollom%
\ \protect \BOthers {.}}{%
McCollom%
\ \protect \BOthers {.}}{%
{\protect \APACyear {2016}}%
}]{%
Mccollom2016}
\APACinsertmetastar {%
Mccollom2016}%
\begin{APACrefauthors}%
McCollom, T\BPBI M.%
, Klein, F.%
, Robbins, M.%
, Moskowitz, B.%
, Berqu{\'o}, T\BPBI S.%
, J{\"o}ns, N.%
\BDBL {}Templeton, A.%
\end{APACrefauthors}%
\unskip\
\newblock
\APACrefYearMonthDay{2016}{}{}.
\newblock
{\BBOQ}\APACrefatitle {Temperature trends for reaction rates, hydrogen
  generation, and partitioning of iron during experimental serpentinization of
  olivine} {Temperature trends for reaction rates, hydrogen generation, and
  partitioning of iron during experimental serpentinization of olivine}.{\BBCQ}
\newblock
\APACjournalVolNumPages{Geochimica et Cosmochimica Acta}{181}{}{175--200}.
\newblock
\begin{APACrefDOI} \doi{10.1016/j.gca.2016.03.002} \end{APACrefDOI}
\PrintBackRefs{\CurrentBib}

\bibitem [\protect \citeauthoryear {%
H.~Menke%
, Jangda%
, Webb%
, Buckman%
\BCBL {}\ \BBA {} Gough%
}{%
H.~Menke%
\ \protect \BOthers {.}}{%
{\protect \APACyear {2025}}%
}]{%
Menke2025Dataset}
\APACinsertmetastar {%
Menke2025Dataset}%
\begin{APACrefauthors}%
Menke, H.%
, Jangda, Z.%
, Webb, M.%
, Buckman, J.%
\BCBL {}\ \BBA {} Gough, A.%
\end{APACrefauthors}%
\unskip\
\newblock
\APACrefYearMonthDay{2025}{}{}.
\newblock
\APACrefbtitle {Hydrogen Generation in an Ultramafic Rock with Micro-CT and
  {SEM-BEX}.} {Hydrogen generation in an ultramafic rock with micro-ct and
  {SEM-BEX}.}
\newblock
\APAChowpublished {Zenodo}.
\newblock
\APACrefnote{[Data set]}
\newblock
\begin{APACrefDOI} \doi{10.5281/zenodo.17989266} \end{APACrefDOI}
\PrintBackRefs{\CurrentBib}

\bibitem [\protect \citeauthoryear {%
H\BPBI P.~Menke%
, Bijeljic%
, Andrew%
\BCBL {}\ \BBA {} Blunt%
}{%
H\BPBI P.~Menke%
\ \protect \BOthers {.}}{%
{\protect \APACyear {2015}}%
}]{%
Menke2015}
\APACinsertmetastar {%
Menke2015}%
\begin{APACrefauthors}%
Menke, H\BPBI P.%
, Bijeljic, B.%
, Andrew, M\BPBI G.%
\BCBL {}\ \BBA {} Blunt, M\BPBI J.%
\end{APACrefauthors}%
\unskip\
\newblock
\APACrefYearMonthDay{2015}{}{}.
\newblock
{\BBOQ}\APACrefatitle {Dynamic three-dimensional pore-scale imaging of reaction
  in a carbonate at reservoir conditions} {Dynamic three-dimensional pore-scale
  imaging of reaction in a carbonate at reservoir conditions}.{\BBCQ}
\newblock
\APACjournalVolNumPages{Environmental Science \&
  Technology}{49}{7}{4407--4414}.
\newblock
\begin{APACrefDOI} \doi{10.1021/es505789f} \end{APACrefDOI}
\PrintBackRefs{\CurrentBib}

\bibitem [\protect \citeauthoryear {%
Miller%
\ \protect \BOthers {.}}{%
Miller%
\ \protect \BOthers {.}}{%
{\protect \APACyear {2016}}%
}]{%
Miller2016}
\APACinsertmetastar {%
Miller2016}%
\begin{APACrefauthors}%
Miller, H\BPBI M.%
, Matter, J\BPBI M.%
, Kelemen, P.%
, Ellison, E\BPBI T.%
, Conrad, M\BPBI E.%
, Fierer, N.%
\BDBL {}Templeton, A\BPBI S.%
\end{APACrefauthors}%
\unskip\
\newblock
\APACrefYearMonthDay{2016}{}{}.
\newblock
{\BBOQ}\APACrefatitle {Modern water/rock reactions in {Oman} hyperalkaline
  peridotite aquifers and implications for microbial habitability} {Modern
  water/rock reactions in {Oman} hyperalkaline peridotite aquifers and
  implications for microbial habitability}.{\BBCQ}
\newblock
\APACjournalVolNumPages{Geochimica et Cosmochimica Acta}{179}{}{217--241}.
\newblock
\begin{APACrefDOI} \doi{10.1016/j.gca.2016.01.033} \end{APACrefDOI}
\PrintBackRefs{\CurrentBib}

\bibitem [\protect \citeauthoryear {%
Miocic%
\ \protect \BOthers {.}}{%
Miocic%
\ \protect \BOthers {.}}{%
{\protect \APACyear {2023}}%
}]{%
Miocic2023}
\APACinsertmetastar {%
Miocic2023}%
\begin{APACrefauthors}%
Miocic, J.%
, Heinemann, N.%
, Edlmann, K.%
, Scafidi, J.%
, Molaei, F.%
\BCBL {}\ \BBA {} Alcalde, J.%
\end{APACrefauthors}%
\unskip\
\newblock
\APACrefYearMonthDay{2023}{}{}.
\newblock
{\BBOQ}\APACrefatitle {Underground hydrogen storage: a review} {Underground
  hydrogen storage: a review}.{\BBCQ}
\newblock
\APACjournalVolNumPages{Geological Society, London, Special
  Publications}{528}{1}{73--86}.
\newblock
\begin{APACrefDOI} \doi{10.1144/sp528-2022-88} \end{APACrefDOI}
\PrintBackRefs{\CurrentBib}

\bibitem [\protect \citeauthoryear {%
Neubeck%
\ \protect \BOthers {.}}{%
Neubeck%
\ \protect \BOthers {.}}{%
{\protect \APACyear {2014}}%
}]{%
Neubeck2014}
\APACinsertmetastar {%
Neubeck2014}%
\begin{APACrefauthors}%
Neubeck, A.%
, Duc, N\BPBI T.%
, Hellevang, H.%
, Oze, C.%
, Bastviken, D.%
, Bacsik, Z.%
\BCBL {}\ \BBA {} Holm, N\BPBI G.%
\end{APACrefauthors}%
\unskip\
\newblock
\APACrefYearMonthDay{2014}{}{}.
\newblock
{\BBOQ}\APACrefatitle {Olivine alteration and {H$_2$} production in
  carbonate-rich, low temperature aqueous environments} {Olivine alteration and
  {H$_2$} production in carbonate-rich, low temperature aqueous
  environments}.{\BBCQ}
\newblock
\APACjournalVolNumPages{Planetary and Space Science}{96}{}{51--61}.
\newblock
\begin{APACrefDOI} \doi{10.1016/j.pss.2014.02.014} \end{APACrefDOI}
\PrintBackRefs{\CurrentBib}

\bibitem [\protect \citeauthoryear {%
Noiriel%
\ \BBA {} Renard%
}{%
Noiriel%
\ \BBA {} Renard%
}{%
{\protect \APACyear {2022}}%
}]{%
NoirielRenard2022}
\APACinsertmetastar {%
NoirielRenard2022}%
\begin{APACrefauthors}%
Noiriel, C.%
\BCBT {}\ \BBA {} Renard, F.%
\end{APACrefauthors}%
\unskip\
\newblock
\APACrefYearMonthDay{2022}{}{}.
\newblock
{\BBOQ}\APACrefatitle {Four-dimensional {X-ray} micro-tomography imaging of
  dynamic processes in geosciences} {Four-dimensional {X-ray} micro-tomography
  imaging of dynamic processes in geosciences}.{\BBCQ}
\newblock
\APACjournalVolNumPages{Comptes Rendus. G{\'e}oscience}{354}{G2}{255--280}.
\newblock
\begin{APACrefDOI} \doi{10.5802/crgeos.137} \end{APACrefDOI}
\PrintBackRefs{\CurrentBib}

\bibitem [\protect \citeauthoryear {%
Noiriel%
\ \BBA {} Soulaine%
}{%
Noiriel%
\ \BBA {} Soulaine%
}{%
{\protect \APACyear {2021}}%
}]{%
NoirielSoulaine2021}
\APACinsertmetastar {%
NoirielSoulaine2021}%
\begin{APACrefauthors}%
Noiriel, C.%
\BCBT {}\ \BBA {} Soulaine, C.%
\end{APACrefauthors}%
\unskip\
\newblock
\APACrefYearMonthDay{2021}{}{}.
\newblock
{\BBOQ}\APACrefatitle {Pore-scale imaging and modelling of reactive flow in
  evolving porous media: Tracking the dynamics of the fluid--rock interface}
  {Pore-scale imaging and modelling of reactive flow in evolving porous media:
  Tracking the dynamics of the fluid--rock interface}.{\BBCQ}
\newblock
\APACjournalVolNumPages{Transport in Porous Media}{140}{1}{181--213}.
\newblock
\begin{APACrefDOI} \doi{10.1007/s11242-021-01613-2} \end{APACrefDOI}
\PrintBackRefs{\CurrentBib}

\bibitem [\protect \citeauthoryear {%
Osselin%
\ \protect \BOthers {.}}{%
Osselin%
\ \protect \BOthers {.}}{%
{\protect \APACyear {2022}}%
}]{%
Osselin2022}
\APACinsertmetastar {%
Osselin2022}%
\begin{APACrefauthors}%
Osselin, F.%
, Soulaine, C.%
, Fauguerolles, C.%
, Gaucher, E\BPBI C.%
, Scaillet, B.%
\BCBL {}\ \BBA {} Pichavant, M.%
\end{APACrefauthors}%
\unskip\
\newblock
\APACrefYearMonthDay{2022}{}{}.
\newblock
{\BBOQ}\APACrefatitle {Orange hydrogen is the new green} {Orange hydrogen is
  the new green}.{\BBCQ}
\newblock
\APACjournalVolNumPages{Nature Geoscience}{15}{10}{765--769}.
\newblock
\begin{APACrefDOI} \doi{10.1038/s41561-022-01043-9} \end{APACrefDOI}
\PrintBackRefs{\CurrentBib}

\bibitem [\protect \citeauthoryear {%
Rezaei%
\ \protect \BOthers {.}}{%
Rezaei%
\ \protect \BOthers {.}}{%
{\protect \APACyear {2022}}%
}]{%
Rezaei2022}
\APACinsertmetastar {%
Rezaei2022}%
\begin{APACrefauthors}%
Rezaei, A.%
, Hassanpouryouzband, A.%
, Molnar, I.%
, Derikvand, Z.%
, Haszeldine, R\BPBI S.%
\BCBL {}\ \BBA {} Edlmann, K.%
\end{APACrefauthors}%
\unskip\
\newblock
\APACrefYearMonthDay{2022}{}{}.
\newblock
{\BBOQ}\APACrefatitle {Relative permeability of hydrogen and aqueous brines in
  sandstones and carbonates at reservoir conditions} {Relative permeability of
  hydrogen and aqueous brines in sandstones and carbonates at reservoir
  conditions}.{\BBCQ}
\newblock
\APACjournalVolNumPages{Geophysical Research Letters}{49}{12}{e2022GL099433}.
\newblock
\begin{APACrefDOI} \doi{10.1029/2022gl099433} \end{APACrefDOI}
\PrintBackRefs{\CurrentBib}

\bibitem [\protect \citeauthoryear {%
Ross%
, Vega%
, Frout{\'e}%
, Kim%
\BCBL {}\ \BBA {} Kovscek%
}{%
Ross%
\ \protect \BOthers {.}}{%
{\protect \APACyear {2025}}%
}]{%
Ross2025_OlivineFlow}
\APACinsertmetastar {%
Ross2025_OlivineFlow}%
\begin{APACrefauthors}%
Ross, C\BPBI M.%
, Vega, B.%
, Frout{\'e}, L.%
, Kim, T\BHBI W.%
\BCBL {}\ \BBA {} Kovscek, A\BPBI R.%
\end{APACrefauthors}%
\unskip\
\newblock
\APACrefYearMonthDay{2025}{}{}.
\newblock
{\BBOQ}\APACrefatitle {Hydrogen generation and serpentinization of olivine
  under flow conditions} {Hydrogen generation and serpentinization of olivine
  under flow conditions}.{\BBCQ}
\newblock
\APACjournalVolNumPages{Geophysical Research Letters}{52}{6}{e2024GL114016}.
\newblock
\begin{APACrefDOI} \doi{10.1029/2024gl114016} \end{APACrefDOI}
\PrintBackRefs{\CurrentBib}

\bibitem [\protect \citeauthoryear {%
Schrenk%
, Brazelton%
\BCBL {}\ \BBA {} Lang%
}{%
Schrenk%
\ \protect \BOthers {.}}{%
{\protect \APACyear {2013}}%
}]{%
Schrenk2013}
\APACinsertmetastar {%
Schrenk2013}%
\begin{APACrefauthors}%
Schrenk, M\BPBI O.%
, Brazelton, W\BPBI J.%
\BCBL {}\ \BBA {} Lang, S\BPBI Q.%
\end{APACrefauthors}%
\unskip\
\newblock
\APACrefYearMonthDay{2013}{}{}.
\newblock
{\BBOQ}\APACrefatitle {Serpentinization, carbon, and deep life}
  {Serpentinization, carbon, and deep life}.{\BBCQ}
\newblock
\APACjournalVolNumPages{Reviews in Mineralogy and
  Geochemistry}{75}{1}{575--606}.
\newblock
\begin{APACrefDOI} \doi{10.2138/rmg.2013.75.18} \end{APACrefDOI}
\PrintBackRefs{\CurrentBib}

\bibitem [\protect \citeauthoryear {%
Shabdirova%
, Silva%
, Isah%
\BCBL {}\ \BBA {} Okoroafor%
}{%
Shabdirova%
\ \protect \BOthers {.}}{%
{\protect \APACyear {2027}}%
}]{%
Shabdirova2027}
\APACinsertmetastar {%
Shabdirova2027}%
\begin{APACrefauthors}%
Shabdirova, A.%
, Silva, H\BPBI G.%
, Isah, A.%
\BCBL {}\ \BBA {} Okoroafor, E\BPBI R.%
\end{APACrefauthors}%
\unskip\
\newblock
\APACrefYearMonthDay{2027}{}{}.
\newblock
{\BBOQ}\APACrefatitle {Low-temperature hydrogen generation from fractured
  ultramafic rock under dynamic flow: implications for engineered geological
  hydrogen systems} {Low-temperature hydrogen generation from fractured
  ultramafic rock under dynamic flow: implications for engineered geological
  hydrogen systems}.{\BBCQ}
\newblock
\APACjournalVolNumPages{Fuel}{428}{}{140187}.
\newblock
\APACrefnote{Part B}
\newblock
\begin{APACrefDOI} \doi{10.1016/j.fuel.2026.140187} \end{APACrefDOI}
\PrintBackRefs{\CurrentBib}

\bibitem [\protect \citeauthoryear {%
Thaysen%
\ \protect \BOthers {.}}{%
Thaysen%
\ \protect \BOthers {.}}{%
{\protect \APACyear {2023}}%
}]{%
Thaysen2023}
\APACinsertmetastar {%
Thaysen2023}%
\begin{APACrefauthors}%
Thaysen, E\BPBI M.%
, Butler, I\BPBI B.%
, Hassanpouryouzband, A.%
, Freitas, D.%
, Alvarez-Borges, F.%
, Krevor, S.%
\BDBL {}Edlmann, K.%
\end{APACrefauthors}%
\unskip\
\newblock
\APACrefYearMonthDay{2023}{}{}.
\newblock
{\BBOQ}\APACrefatitle {Pore-scale imaging of hydrogen displacement and trapping
  in porous media} {Pore-scale imaging of hydrogen displacement and trapping in
  porous media}.{\BBCQ}
\newblock
\APACjournalVolNumPages{International Journal of Hydrogen
  Energy}{48}{8}{3091--3106}.
\newblock
\begin{APACrefDOI} \doi{10.1016/j.ijhydene.2022.10.153} \end{APACrefDOI}
\PrintBackRefs{\CurrentBib}

\bibitem [\protect \citeauthoryear {%
Wang%
, Watanabe%
, Okamoto%
, Nakamura%
\BCBL {}\ \BBA {} Komai%
}{%
Wang%
\ \protect \BOthers {.}}{%
{\protect \APACyear {2019}}%
}]{%
Wang2019}
\APACinsertmetastar {%
Wang2019}%
\begin{APACrefauthors}%
Wang, J.%
, Watanabe, N.%
, Okamoto, A.%
, Nakamura, K.%
\BCBL {}\ \BBA {} Komai, T.%
\end{APACrefauthors}%
\unskip\
\newblock
\APACrefYearMonthDay{2019}{}{}.
\newblock
{\BBOQ}\APACrefatitle {Enhanced hydrogen production with carbon storage by
  olivine alteration in {CO$_2$}-rich hydrothermal environments} {Enhanced
  hydrogen production with carbon storage by olivine alteration in
  {CO$_2$}-rich hydrothermal environments}.{\BBCQ}
\newblock
\APACjournalVolNumPages{Journal of CO2 Utilization}{30}{}{205--213}.
\newblock
\begin{APACrefDOI} \doi{10.1016/j.jcou.2019.02.008} \end{APACrefDOI}
\PrintBackRefs{\CurrentBib}

\bibitem [\protect \citeauthoryear {%
Wong%
, Cascone%
, Giovannelli%
\BCBL {}\ \BBA {} Vitale~Brovarone%
}{%
Wong%
\ \protect \BOthers {.}}{%
{\protect \APACyear {2026}}%
}]{%
Wong2026}
\APACinsertmetastar {%
Wong2026}%
\begin{APACrefauthors}%
Wong, K.%
, Cascone, M.%
, Giovannelli, D.%
\BCBL {}\ \BBA {} Vitale~Brovarone, A.%
\end{APACrefauthors}%
\unskip\
\newblock
\APACrefYearMonthDay{2026}{}{}.
\newblock
{\BBOQ}\APACrefatitle {Formation and fluxes of natural hydrogen in the crust
  and upper mantle} {Formation and fluxes of natural hydrogen in the crust and
  upper mantle}.{\BBCQ}
\newblock
\APACjournalVolNumPages{Earth-Science Reviews}{275}{}{105400}.
\newblock
\begin{APACrefDOI} \doi{10.1016/j.earscirev.2026.105400} \end{APACrefDOI}
\PrintBackRefs{\CurrentBib}

\bibitem [\protect \citeauthoryear {%
Wu%
\ \protect \BOthers {.}}{%
Wu%
\ \protect \BOthers {.}}{%
{\protect \APACyear {2026}}%
}]{%
Wu2026}
\APACinsertmetastar {%
Wu2026}%
\begin{APACrefauthors}%
Wu, Z.%
, Ge, X.%
, Li, Z.%
, Neil, C\BPBI W.%
, Wang, S.%
\BCBL {}\ \BBA {} Li, W.%
\end{APACrefauthors}%
\unskip\
\newblock
\APACrefYearMonthDay{2026}{}{}.
\newblock
{\BBOQ}\APACrefatitle {Experimental Parameterization of Geological Hydrogen
  Production via Stimulated Serpentinization} {Experimental parameterization of
  geological hydrogen production via stimulated serpentinization}.{\BBCQ}
\newblock
\APACjournalVolNumPages{Rock Mechanics and Rock Engineering}{}{}{}.
\newblock
\APACrefnote{Advance online publication}
\newblock
\begin{APACrefDOI} \doi{10.1007/s00603-026-05756-w} \end{APACrefDOI}
\PrintBackRefs{\CurrentBib}

\bibitem [\protect \citeauthoryear {%
Xia%
, Espinoza%
, Steefel%
, Manga%
\BCBL {}\ \BBA {} Hu%
}{%
Xia%
\ \protect \BOthers {.}}{%
{\protect \APACyear {2026}}%
}]{%
Xia2026}
\APACinsertmetastar {%
Xia2026}%
\begin{APACrefauthors}%
Xia, Z.%
, Espinoza, D\BPBI N.%
, Steefel, C.%
, Manga, M.%
\BCBL {}\ \BBA {} Hu, M.%
\end{APACrefauthors}%
\unskip\
\newblock
\APACrefYearMonthDay{2026}{}{}.
\newblock
{\BBOQ}\APACrefatitle {Spatiotemporal monitoring of serpentinization in
  peridotite: mineral replacement, volume change and hydrogen production}
  {Spatiotemporal monitoring of serpentinization in peridotite: mineral
  replacement, volume change and hydrogen production}.{\BBCQ}
\newblock
\APACjournalVolNumPages{Geochimica et Cosmochimica Acta}{425}{}{116--130}.
\newblock
\begin{APACrefDOI} \doi{10.1016/j.gca.2026.04.029} \end{APACrefDOI}
\PrintBackRefs{\CurrentBib}

\bibitem [\protect \citeauthoryear {%
Yekta%
, Manceau%
, Gaboreau%
, Pichavant%
\BCBL {}\ \BBA {} Audigane%
}{%
Yekta%
\ \protect \BOthers {.}}{%
{\protect \APACyear {2018}}%
}]{%
Yekta2018}
\APACinsertmetastar {%
Yekta2018}%
\begin{APACrefauthors}%
Yekta, A\BPBI E.%
, Manceau, J\BHBI C.%
, Gaboreau, S.%
, Pichavant, M.%
\BCBL {}\ \BBA {} Audigane, P.%
\end{APACrefauthors}%
\unskip\
\newblock
\APACrefYearMonthDay{2018}{}{}.
\newblock
{\BBOQ}\APACrefatitle {Determination of hydrogen--water relative permeability
  and capillary pressure in sandstone: application to underground hydrogen
  injection in sedimentary formations} {Determination of hydrogen--water
  relative permeability and capillary pressure in sandstone: application to
  underground hydrogen injection in sedimentary formations}.{\BBCQ}
\newblock
\APACjournalVolNumPages{Transport in Porous Media}{122}{2}{333--356}.
\newblock
\begin{APACrefDOI} \doi{10.1007/s11242-018-1004-7} \end{APACrefDOI}
\PrintBackRefs{\CurrentBib}

\bibitem [\protect \citeauthoryear {%
Zgonnik%
}{%
Zgonnik%
}{%
{\protect \APACyear {2020}}%
}]{%
Zgonnik2020}
\APACinsertmetastar {%
Zgonnik2020}%
\begin{APACrefauthors}%
Zgonnik, V.%
\end{APACrefauthors}%
\unskip\
\newblock
\APACrefYearMonthDay{2020}{}{}.
\newblock
{\BBOQ}\APACrefatitle {The occurrence and geoscience of natural hydrogen: A
  comprehensive review} {The occurrence and geoscience of natural hydrogen: A
  comprehensive review}.{\BBCQ}
\newblock
\APACjournalVolNumPages{Earth-Science Reviews}{203}{}{103140}.
\newblock
\begin{APACrefDOI} \doi{10.1016/j.earscirev.2020.103140} \end{APACrefDOI}
\PrintBackRefs{\CurrentBib}

\end{thebibliography}


\begin{thebibliography}{}

\bibitem [\protect \citeauthoryear {%
Buades%
, Coll%
\BCBL {}\ \BBA {} Morel%
}{%
Buades%
\ \protect \BOthers {.}}{%
{\protect \APACyear {2005}}%
}]{%
Buades2005}
\APACinsertmetastar {%
Buades2005}%
\begin{APACrefauthors}%
Buades, A.%
, Coll, B.%
\BCBL {}\ \BBA {} Morel, J\BHBI M.%
\end{APACrefauthors}%
\unskip\
\newblock
\APACrefYearMonthDay{2005}{}{}.
\newblock
{\BBOQ}\APACrefatitle {A non-local algorithm for image denoising} {A non-local
  algorithm for image denoising}.{\BBCQ}
\newblock
\BIn{} \APACrefbtitle {2005 {IEEE Computer Society} Conference on Computer
  Vision and Pattern Recognition ({CVPR}'05)} {2005 {IEEE Computer Society}
  conference on computer vision and pattern recognition ({CVPR}'05)}\ (\BVOL~2,
  \BPGS\ 60--65).
\newblock
\begin{APACrefDOI} \doi{10.1109/cvpr.2005.38} \end{APACrefDOI}
\PrintBackRefs{\CurrentBib}

\bibitem [\protect \citeauthoryear {%
{IAPWS}%
}{%
{IAPWS}%
}{%
{\protect \APACyear {1992}}%
}]{%
IAPWS1992Saturation}
\APACinsertmetastar {%
IAPWS1992Saturation}%
\begin{APACrefauthors}%
{IAPWS}.%
\end{APACrefauthors}%
\unskip\
\newblock
\APACrefYearMonthDay{1992}{{\APACmonth{09}}}{}.
\newblock
\APACrefbtitle {Revised Supplementary Release on Saturation Properties of
  Ordinary Water Substance} {Revised supplementary release on saturation
  properties of ordinary water substance}\ \APACbVolEdTR {}{Supplementary
  Release\ \BNUM\ SR1-86(1992)}.
\newblock
\APACaddressInstitution{}{International Association for the Properties of Water
  and Steam}.
\newblock
\begin{APACrefURL}
  \url{https://iapws.org/technical-guidance/release/Supp-sat.download}
  \end{APACrefURL}
\PrintBackRefs{\CurrentBib}

\bibitem [\protect \citeauthoryear {%
{IAPWS}%
}{%
{IAPWS}%
}{%
{\protect \APACyear {2004}}%
}]{%
IAPWS2004Henry}
\APACinsertmetastar {%
IAPWS2004Henry}%
\begin{APACrefauthors}%
{IAPWS}.%
\end{APACrefauthors}%
\unskip\
\newblock
\APACrefYearMonthDay{2004}{{\APACmonth{09}}}{}.
\newblock
\APACrefbtitle {Guideline on the {Henry's} Constant and Vapor-Liquid
  Distribution Constant for Gases in {H2O} and {D2O} at High Temperatures}
  {Guideline on the {Henry's} constant and vapor-liquid distribution constant
  for gases in {H2O} and {D2O} at high temperatures}\ \APACbVolEdTR
  {}{Guideline\ \BNUM\ G7-04}.
\newblock
\APACaddressInstitution{}{International Association for the Properties of Water
  and Steam}.
\newblock
\begin{APACrefURL} \url{https://iapws.org/documents/release/HenGuide.download}
  \end{APACrefURL}
\PrintBackRefs{\CurrentBib}

\bibitem [\protect \citeauthoryear {%
Lanczos%
}{%
Lanczos%
}{%
{\protect \APACyear {1956}}%
}]{%
Lanczos1956}
\APACinsertmetastar {%
Lanczos1956}%
\begin{APACrefauthors}%
Lanczos, C.%
\end{APACrefauthors}%
\unskip\
\newblock
\APACrefYear{1956}.
\newblock
\APACrefbtitle {Applied Analysis} {Applied analysis}.
\newblock
\APACaddressPublisher{Englewood Cliffs, NJ}{Prentice-Hall}.
\PrintBackRefs{\CurrentBib}

\end{thebibliography}
\end{document}

\title{Supporting Information for "A First Look at Hydrogen Generation in an Ultramafic Rock with Micro-CT and SEM-BEX"}

\authors{Menke, Hannah P.\affil{1,3}*; Jangda, Zaid Z.\affil{2,3}; Jones, CJ\affil{1,3}; Webb, Max\affil{1,3}; Buckman, Jim\affil{1}; Gough, Amy\affil{1,3}}
\affiliation{1}{Institute of GeoEnergy Engineering, Heriot-Watt University, Edinburgh, UK}
\affiliation{2}{The Lyell Centre, Heriot-Watt University, Edinburgh, UK}
\affiliation{3}{Subsurface Energy Transition \& Innovation Centre, Heriot-Watt University, Edinburgh, UK}

\noindent\textbf{Contents of this file}
\begin{enumerate}
\item Text S1: imaging and microscopy methods; Text S2: companion reactor assay
\item Figures S1--S4
\item Tables S1--S6
\end{enumerate}
\noindent\textbf{Additional Supporting Information (Files uploaded separately)}
\begin{enumerate}
\item None.
\end{enumerate}

\noindent\textbf{Introduction}
This Supporting Information provides expanded methods, additional figures and tables, and hydrogen mass-balance calculations.

\noindent\textbf{Text S1. Expanded Materials and Methods}
\section{Sample Selection and Preparation}

The experiment used an ultramafic rock sample selected for its potential for low-temperature serpentinization. The sample was collected from the mantle sequence of the Cyclops Ophiolite located in the Cyclops Mountains of Papua Province, Indonesia (along the northern coast of the island of New Guinea).

\begin{figure}
  \centering
  \includegraphics[width=\linewidth]{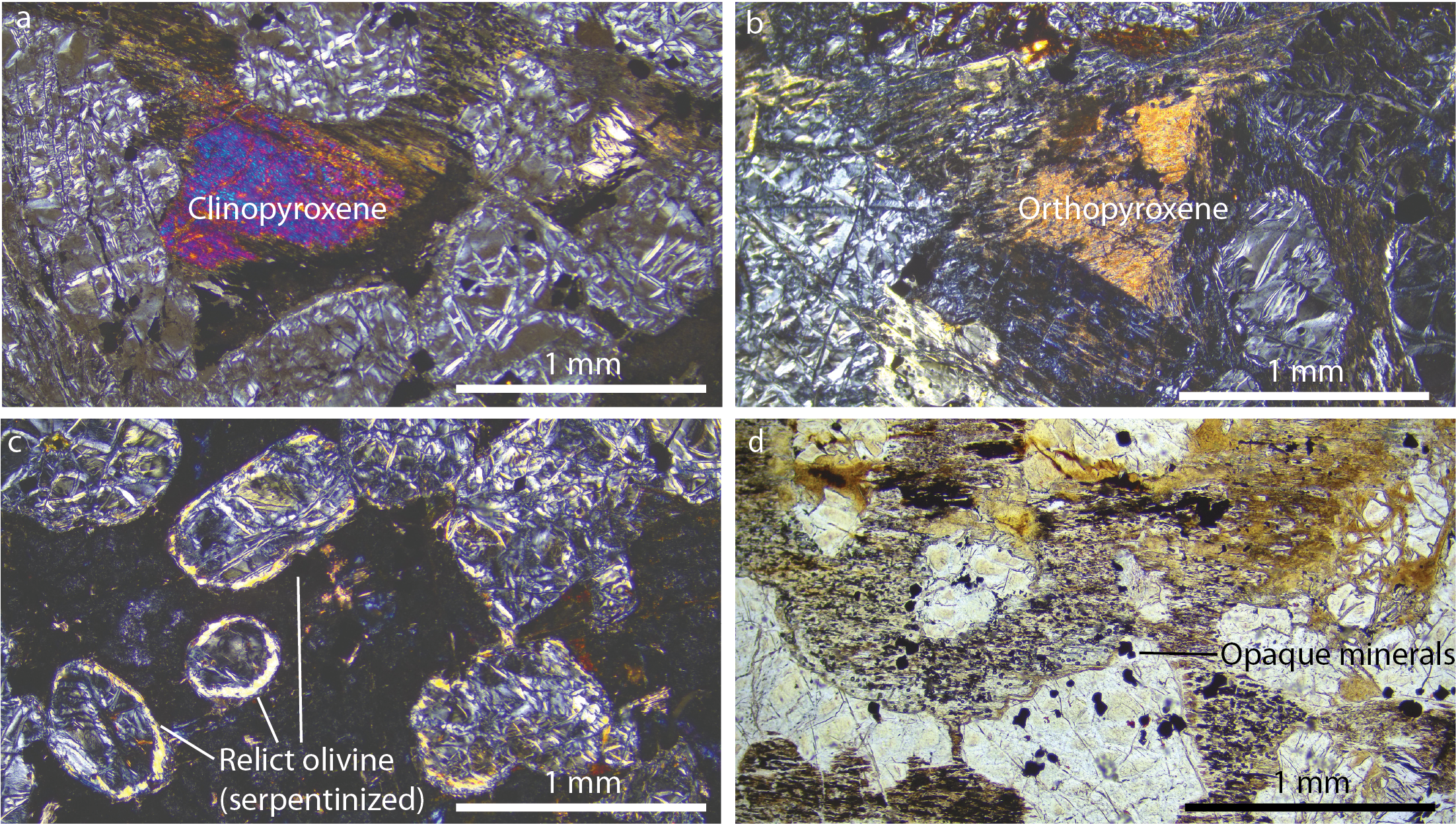}
  \caption{Thin-section photomicrographs of PS23-011 showing grain size and euhedral crystal textures under cross-polarized (a--c) and plane-polarized (d) light. Panels show (a) clinopyroxene, (b) orthopyroxene, (c) olivine relics replaced by serpentine and (d) opaque minerals. Scale bars: 1 mm.}
  \label{fig:petrography}
\end{figure}

Thin-section examination in transmitted (plane- and cross-polarized) and reflected light identifies PS23-011 as a serpentinized lherzolite with an equigranular texture and crystal sizes of 0.5--1 mm (Figure~\ref{fig:petrography}). The estimated proportions are approximately 50\% olivine-derived domains, 10\% clinopyroxene (diopside), 30\% orthopyroxene (enstatite) and 10\% opaque minerals. The olivine is almost entirely replaced by serpentine. Clinopyroxene is partially altered but not serpentinized, whereas orthopyroxene shows local serpentinization. The opaque minerals appear light grey in reflected light and are tentatively identified as magnetite. Serpentine occurs as rims around olivine relics and within fractures. Pyroxene oikocrysts locally enclose relict olivine chadocrysts.

The rock was initially broken using a steel geological hammer and subsequently pulverised using a ceramic pestle and mortar. Grain-size separation was performed using steel sieves. The grains were not washed, rinsed or otherwise cleaned after preparation.

The prepared rock was dry sieved to isolate the coarse fraction, retaining only grains between 0.6-1 mm. This grain size was selected to ensure a resolvable pore network for 3D imaging while maintaining reactive surface area.

\section{Experimental Setup and Protocol}

The imaging experiment was conducted at 100$^\circ$C. The pore space was saturated with KI-doped brine at a pore pressure of 4~bar, while the core holder was loaded within a Hassler-type sleeve under an external confining pressure of 10~bar. Unless otherwise stated, imaging pressures refer to pore pressure. The pore pressure exceeds the saturation vapour pressure of water at 100$^\circ$C, suppressing bulk boiling. Water vapour can nevertheless contribute to a gas phase containing non-condensable species.

\begin{figure}
  \centering
  \includegraphics[width=0.7\textwidth]{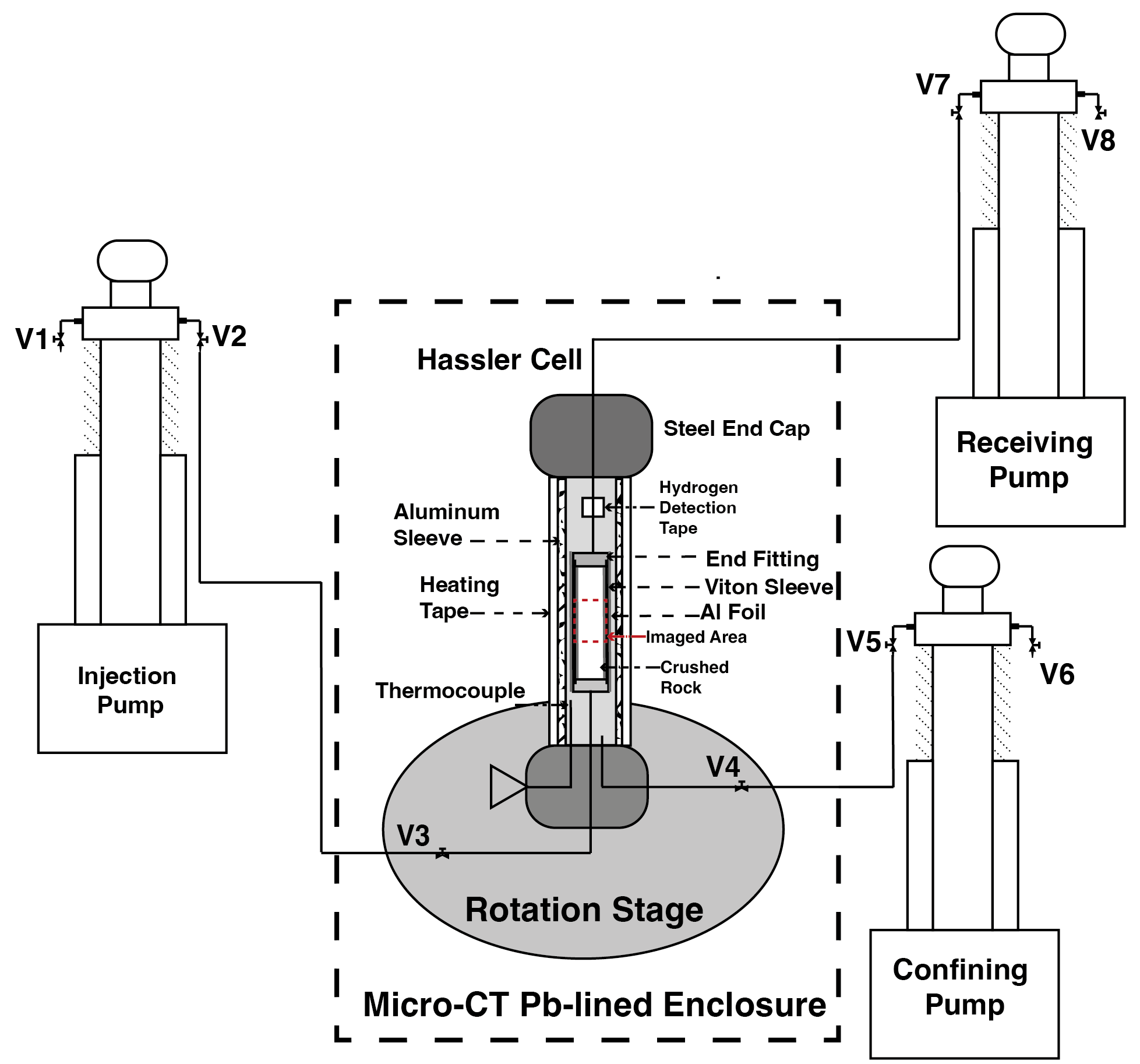}
  \caption{Experimental set-up for in situ micro-CT imaging of H$_2$ generation and gas exsolution. A Hassler-type core holder containing crushed ultramafic rock is mounted on the rotation stage inside a Pb-lined micro-CT enclosure. The sample is confined by a Viton sleeve and aluminium foil within an aluminium outer sleeve, with steel end caps and end fittings connecting to the flow lines. Heating tape and a thermocouple control and monitor the sample temperature, and H$_2$-indicator tape is placed around the core to detect any leakage. Pore fluid is injected from the injection pump, while the receiving pump maintains pressure and collects effluent. Confining pressure is applied via the confining pump. The red dashed box indicates the imaged region within the core.}
  \label{fig:flowapparatus}
\end{figure}

The granular sample was loaded into a custom-built X-ray transparent core holder designed for high-pressure and high-temperature flow experiments. A strip of H$_2$ indicator was attached inside the confining sleeve of the core holder to detect the presence of H$_2$. The rig was mounted within the X-ray micro-tomography (micro-CT) scanner to facilitate \emph{in situ} imaging (Fig. \ref{fig:flowapparatus}).

The initial saturation protocol was designed to reduce trapped air and establish liquid saturation prior to heating. The sample was flushed with CO$_2$ at ambient conditions and confined at 10 bar. The 5~wt\% potassium iodide (KI) brine was degassed under vacuum for 36 h before injection. The sample was then flooded with this brine at 4~bar pore pressure. KI enhanced X-ray attenuation of the liquid, providing contrast between grains, brine and gas. The reported 1000 pore-volume brine flush was intended to dissolve and displace the initial CO$_2$; residual dissolved carbon was not measured.

Following the brine saturation phase, flow was halted (static conditions with no net through-flow), and the temperature of the sample was raised to 100$^\circ$C to initiate the hydrothermal reaction. During this static stage, pore pressure is maintained at 4~bar, so any volumetric expansion associated with gas exsolution was accommodated by small displacements of brine into the pump rather than by a significant increase in pore pressure.

After the imaging experiment, the receiving pump was isolated from the core and its contents discharged into a large Pyrex vessel containing a portable electrochemical hydrogen detector (RAE ToxiRAE Pro PGM-1860). H$_2$ in the pump gas effluent was monitored during controlled depressurisation. The reacted sample was then removed from the core holder, and the H$_2$-indicator strip was retrieved from the confining sleeve and inspected.

\section{Micro-CT Image Acquisition}
Time-resolved 3D images were acquired using an Rx EasyTom micro-CT scanner with a voxel size of 6.277 $\mu$m. Images were acquired approximately every 50 min during the early period; the first resolved gas is reported at approximately 8 h. The sample was subsequently imaged at 23 and 30 h. Each acquisition lasted approximately 50 min and used 800 projections, 9.9 W and 2 frames/s. The source setting was recorded as 100 keV. The nominal scan times describe sampled states, not instantaneous measurements of nucleation, and features below the spatial resolution may remain unresolved.

\section{Micro-CT Image Analysis}

\begin{figure}
  \centering
  \includegraphics[width=\linewidth]{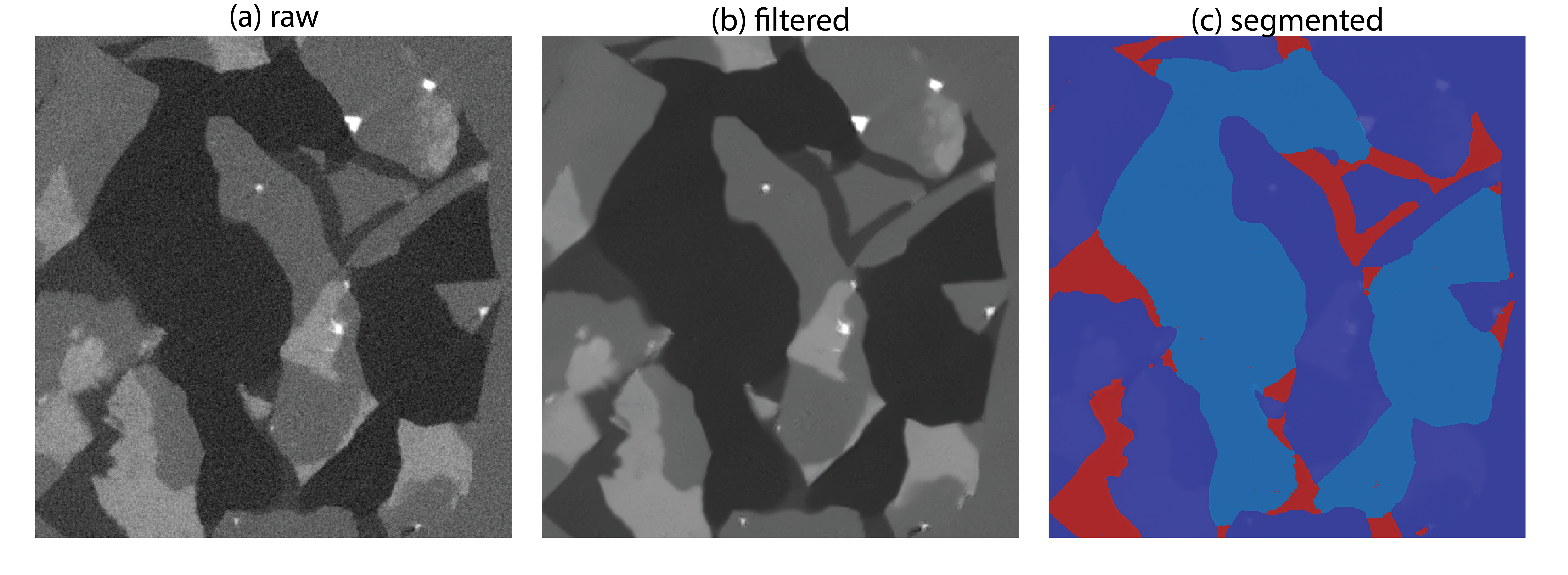}
  \caption{Representative 2D slice showing the micro-CT image processing workflow. (a) Raw greyscale image, (b) the same slice after non-local means filtering to reduce high-frequency noise while preserving phase boundaries, and (c) final three-phase segmentation into solid grains (dark blue), brine-filled pore space (red) and gas voxels (light blue).}
  \label{fig:ImageProcessing}
\end{figure}

To quantify the evolution of the gas phase, all time-lapse micro-CT volumes were processed using a consistent image-analysis workflow in the image processing software Avizo 2025.1 (Thermo Fisher Scientific). First, each scan was rigidly registered to the initial reference scan by maximising mutual information between grey-scale volumes and resampled onto the reference grid using a Lanczos kernel \citep{Lanczos1956} to preserve edge sharpness. Second, we applied a non-local means filter to reduce high-frequency noise while retaining phase boundaries in the greyscale images \citep{Buades2005}. Third, the solid, brine and gas phases were segmented in the Avizo Segmentation Editor using simple global grey-level thresholding, guided by the histogram and visual inspection of representative slices. Conservative thresholds were chosen to separate solid from pore space, and within the pore space voxels with the lowest attenuation were classified as gas. Small isolated gas clusters below a few voxels were removed as likely noise using basic morphological operations. A representative 2D slice illustrating the raw, filtered and segmented images is shown in Figure~\ref{fig:ImageProcessing}. For each time step we then calculated bulk gas saturation as the ratio of gas-filled voxels to pore-space voxels. Visual inspection of representative slices and 3D renderings was used to check the segmentation quality; we therefore regard the resulting saturation values as approximate but internally consistent across the time series.

We assessed sensitivity to the gas--brine threshold by inspecting alternative segmentations consistent with the grey-level histograms and visible phase boundaries. These alternative segmentations yielded very similar temporal trends in bulk $S_g$ and connectivity, and in particular preserved the three-stage evolution (no resolvable gas, rapid increase, slower late-time growth).

Gas ganglia statistics were obtained from the segmented gas phase using Avizo's \textit{Label Analysis} module. Connected-component labelling was applied to the binary gas mask to identify individual gas clusters (ganglia), and for each cluster we extracted its volume. From these we computed the distribution of ganglia sizes, the number density of clusters, and the volume fraction of the largest cluster $f_{\max}$ at each time step. 

The largest-cluster fraction is the proportion of segmented gas contained in the largest connected cluster and is reported only for scans containing gas.

\section{SEM and BEX analysis}

Pre- and post-reaction grains for electron microscopy were prepared from the same bulk material as the micro-CT sample. Sub-samples were gently dry-sieved to remove dust, sprinkled onto aluminium stubs, and immobilised using conductive adhesive carbon tabs (Agar Scientific). Grains were dried without rinsing to remove KI; some surface features may therefore reflect salt precipitation during drying. Post-experiment SEM images show discrete platy features and roughened patches consistent with local surface alteration.

The mounts were imaged using an FEI Quanta FEG 650 Scanning Electron Microscope (SEM) operated in low-vacuum mode. Backscattered electron (BSE) images were acquired at magnifications of hundreds to several thousand times to capture both grain-scale morphology and fine-scale surface textures. For each mount, we collected combined backscatter and energy-dispersive X-ray (BEX) data using an Oxford Instruments Unity detector in combination with an X-Max$_N$ 150 mm Energy Dispersive X-ray (EDX) detector.  These generate colour-composite elemental maps by assigning selected element signals (e.g., Mg, Al, Si, Ca, Fe) to different colour channels along with quantitative elemental analysis. These BEX composites were used to visualise spatial variations in surface composition across many grains using the Oxford Instruments AZtec 6.2\texttrademark\ software. In addition, five spot elemental analyses were acquired on each mount, targeting representative areas on the larger grains, and the resulting ranges in composition are reported in Table~\ref{tab:xrf_bex_aligned}.  No matrix-specific corrections were applied beyond the standard instrument routines, so the spot analyses are interpreted qualitatively and in conjunction with the textural information from the BSE and BEX images.

\begin{figure}
  \centering
  \includegraphics[width=0.7\textwidth]{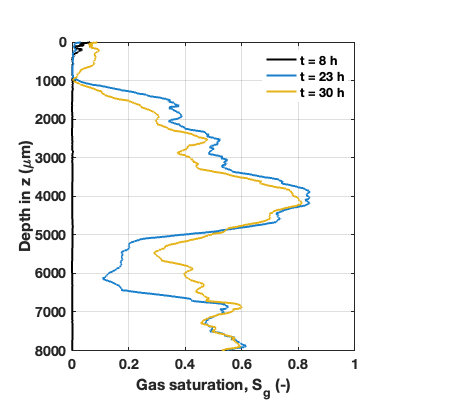}
  \caption{Vertical profiles of gas saturation $S_g$ as a function of depth $z$ along the core axis at 8, 23 and 30~h. Depth increases downward from the top of the imaged sample and is expressed in micrometres; line colours correspond to the times indicated in the legend.}
  \label{fig:zaxisprofile}
\end{figure}

\begin{table}[H]
  \centering
  \caption{Image-derived geometry and gas-occupancy metrics. $\phi$ is porosity, $A$ the reported solid--pore interfacial area, $S_g$ the bulk gas saturation, $\Delta S_g / \Delta t$ the reported interval-average change between scans, and $f_{\max}$ the fraction of segmented gas in the largest cluster. Saturations are rounded; interval slopes retain values supplied in the original analysis and are not H$_2$ production rates.}
  \label{tab:image_metrics}

  \textit{(a) Static micro-CT geometry at $t=0$.}

  \begin{tabular}{lccc}
    \toprule
    $t$ & $\phi$ & $A/V_s$ & $A$ \\
    (h) & (-)    & ($\mathrm{m^2\,m^{-3}}$) & ($\mathrm{m^2}$) \\
    \midrule
    0 & 0.375 & 1.13$\times 10^{4}$ & 4.27$\times 10^{-4}$ \\
    \bottomrule
  \end{tabular}

  \vspace{0.75em}

  \textit{(b) Time-lapse gas saturation and connectivity metrics.}

  \begin{tabular}{lccccc}
    \toprule
    State & $t$ & $S_g$ & $\Delta t$ & $\Delta S_g/\Delta t$ & $f_{\max}$ \\
         & (h) & (-)   & (h)        & ($\mathrm{h^{-1}}$)   & (-) \\
    \midrule
    1 & 0   & 0.00 & --  & --                    & n.a. \\
    2 & 8   & 0.01 & 8   & 8.74$\times 10^{-4}$  & 0.501 \\
    3 & 23  & 0.39 & 15  & 2.56$\times 10^{-2}$  & 0.996 \\
    4 & 30  & 0.43 & 7   & 5.87$\times 10^{-3}$  & 0.964 \\
    \bottomrule
  \end{tabular}

\smallskip
\noindent n.a.: no segmented gas.
\end{table}

\begin{table}[H]
  \centering
  \caption{Elemental composition from bulk XRF and BEX spot analyses before (pre) and after (post) reaction. BEX values give the minimum--maximum range of wt.\,\% across five spot analyses on each mount. Dashes indicate elements not detected in any of the BEX spots.}
  \label{tab:xrf_bex_aligned}
  \begin{tabular}{lccc}
    \toprule
    & \multicolumn{1}{c}{Bulk XRF} & \multicolumn{1}{c}{BEX pre} & \multicolumn{1}{c}{BEX post} \\
    \cmidrule(lr){2-2} \cmidrule(lr){3-3} \cmidrule(lr){4-4}
    Element & wt.\,\% & wt.\,\% (range) & wt.\,\% (range) \\
    \midrule
    O   & --     & 48.10--51.92 & 49.48--51.58 \\
    C   & --     & 5.09--6.64   & --           \\
    Na  & 0.045  & --           & --           \\
    Mg  & 20.127 & 18.88--20.97 & 14.44--21.10 \\
    Al  & 2.213  & 1.44--2.50   & 3.09--6.58   \\
    Si  & 17.524 & 16.38--18.97 & 15.19--17.49 \\
    P   & 0.013  & --           & --           \\
    S   & 0.008  & --           & --           \\
    K   & 0.003  & --           & 0.23--0.23   \\
    Ca  & 1.780  & 0.36--0.36   & 0.26--6.11   \\
    Ti  & 0.081  & 0.06--0.06   & --           \\
    Mn  & 0.108  & 0.13--1.22   & 0.34--0.34   \\
    Fe  & 6.542  & 5.32--6.80   & 5.19--7.29   \\
    Ni  & 0.156  & 0.17--0.42   & 0.20--0.20   \\
    Cr  & 0.691  & 0.11--0.30   & 0.40--0.40   \\
    Sr  & 0.001  & --           & --           \\
    Zn  & 0.018  & --           & --           \\
    Ba  & 0.009  & --           & --           \\
    Zr  & 0.006  & --           & --           \\
    Yb  & --     & --           & 0.90--0.90   \\
    \bottomrule
  \end{tabular}
\end{table}

\section{Early hydrogen inventory and saturation timescale}
The early hydrogen inventory was reconstructed from the first gas-bearing scan at 8 h. The model uses initially negligible hydrogen, constant net generation over 0--8 h, equilibrium gas--water partitioning with gas formation once the brine reached saturation, and no gas transfer across the boundary of the imaged pore volume. It accounts for saturated brine displaced by gas expansion at constant pressure.

The equilibrium dissolved and gas-phase hydrogen concentrations are $C_l=k_Hp_{\mathrm{H_2}}$ and $C_g=p_{\mathrm{H_2}}/(RT)$, respectively. Accounting for hydrogen retained in the liquid and gas, together with hydrogen exported in displaced brine, gives, after saturation,
\begin{equation}
N(t)=C_lV_p(1-S_g)+C_gV_pS_g+C_lV_pS_g=C_lV_p+C_gV_pS_g.
\end{equation}
The mean generation rate over the first interval is $G=N(8\,\mathrm{h})/(8\,\mathrm{h})$, giving
\begin{equation}
t_{\mathrm{sat}}=\frac{C_lV_p}{G}=\frac{8\,\mathrm{h}\,\alpha}{\alpha+S_g(8\,\mathrm{h})},\qquad\alpha=k_HRT.
\end{equation}
Hydrogen partial pressure and pore volume cancel in this expression. The IAPWS pure-water Henry coefficient at 100$^\circ$C, converted using saturated-water density, is $k_H=7.36\times10^{-4}\ \mathrm{mol\,L^{-1}\,bar^{-1}}$, giving $\alpha=0.0228$ \citep{IAPWS2004Henry,IAPWS1992Saturation}. The rounded saturation of 0.01 gives $t_{\mathrm{sat}}=5.6$ h; the saturation of 0.00699 implied by the tabulated first-interval slope gives 6.1 h.

This reconstruction applies to the early interval. Extending the same generation rate to 23 h gives $S_g=0.06$--$0.07$, compared with the observed value of approximately 0.39. The pure-water coefficient is used as a reference approximation for the KI brine.

\clearpage
\noindent\textbf{Text S2. Companion reactor assay and processing}
\section{Experimental conditions and data provenance}
The companion experiment used approximately 200 g of crushed PS23-011 and 430 mL of degassed deionised water in a 600 mL reactor. The rock displaced approximately 70 mL, leaving a nominal 100 mL helium headspace. The reactor was maintained at 100$^\circ$C and nominally 4 bar gauge, with helium supplied by a pressure-control pump through a dip tube to maintain pressure during gas sampling. No liquid samples were collected, and no liquid was observed when the pressure-control pump was emptied at the end of the experiment. Table S6 summarises the conditions of the two experiments.

After loading the rock, the reactor was evacuated and flushed with helium. Degassed deionised water was transferred to the pump under a helium flow and then injected into the reactor. The reactor was flushed again with helium before heating and subsequent pressurisation.

Baseline-corrected median ion signals from nine mass spectra were used to characterise the pressurised reactor time series. Detector settings were consistent between samples. An additional spectrum collected after depressurisation was excluded. Sampling times and elapsed durations are reported in Tables S4 and S5.

\section{Signals and relative sensitivity factors}
Let $I_j$ denote the supplied baseline-corrected median signal at mass-to-charge channel $j$. The main-text reactor figure shows nominal sampled-gas H$_2$ mole fraction and estimated cumulative net H$_2$ production, accounting for dissolved hydrogen and sampling losses. Gas mole fractions were estimated using N$_2$-normalised relative sensitivity factors. Instrument-specific values were 0.435 for H$_2$, 0.175 for He and 1.378 for CO$_2$, determined using gas standards. Literature values of 1.12 for H$_2$O and 0.86 for O$_2$ were retained, with N$_2$ set to 1.00. The standards were BOC N$_2$ Zero (99.998\% purity), H$_2$ N4.5 (99.995\%), CO$_2$ N4.5 (99.995\%) and He CP (99.999\%).

The nominal mole fraction for species $j$ is computed as
\begin{equation}
 \widehat{x}_j=\frac{I_j/\mathrm{RSF}_j}{\sum_k I_k/\mathrm{RSF}_k},
\end{equation}
where $j$ and $k$ index H$_2$, He, H$_2$O, N$_2$, O$_2$ and CO$_2$, assigned to $m/z$ 2, 4, 18, 28, 32 and 44, respectively. Each channel was assigned to the corresponding species. The hydrogen mass balance used dry-gas mole fractions normalised over H$_2$, He, N$_2$, O$_2$ and CO$_2$, excluding the water channel. The transfer calculation conserves ratios among these gases and treats water separately through its saturation vapour pressure. A signal at $m/z=44$ is consistent with CO$_2$ but is not a determination of carbon source.

\section{Hydrogen mass balance}
Each withdrawal filled the evacuated 12.4 mL sampling loop at ambient temperature. The sampling valve was closed after the pressure-control pump had restored the reactor pressure. The mass balance used 0.430 kg of water and a hot headspace volume of approximately 81 mL, accounting for water expansion at 100$^\circ$C. Absolute pressures were calculated by adding 1.01325 bar to the recorded pressures. Sampling-loop temperatures of 20 and 25$^\circ$C were evaluated in the calculation.

The dry-gas hydrogen mole fraction, $y_{\mathrm{H_2,dry}}$, was calculated using the signal-normalisation equation above with the denominator restricted to the five noncondensing gases. The hydrogen partial pressure in the reactor was calculated as $p_{\mathrm{H_2}}=y_{\mathrm{H_2,dry}}[P-p_{\mathrm{sat}}(T_r)]$, where $P$ is absolute pressure, $T_r=373.15$ K and $p_{\mathrm{sat}}$ is the saturation vapour pressure of water. The headspace inventory was calculated using $n_g=p_{\mathrm{H_2}}V_g/(RT_r)$, with $V_g=0.0813$ L and $R=0.0831446$ L bar mol$^{-1}$ K$^{-1}$; pressures are expressed in bar. Water density and saturation vapour pressure were calculated using the IAPWS expressions \citep{IAPWS2004Henry,IAPWS1992Saturation}.

Hydrogen removed in each sampling withdrawal was calculated as $n_i=y_{\mathrm{H_2,dry},i}[P_i-p_{\mathrm{sat}}(T_l)]V_l/(RT_l)$, with $V_l=0.0124$ L and $T_l=293.15$ or 298.15 K. The gas balance for each sample included all preceding withdrawals, including those without usable spectra. For withdrawals without usable spectra, the dry-gas hydrogen mole fraction was interpolated linearly in time between the adjacent successful samples. Water-vapour partial pressures were calculated at saturation at the reactor and sampling-loop temperatures.

Dissolved hydrogen was calculated at gas--liquid equilibrium using $n_{\mathrm{aq}}=k_H(T_r)m_w p_{\mathrm{H_2}}$, where $k_H$ is expressed in mol kg$^{-1}$ bar$^{-1}$, $m_w=0.430$ kg and $p_{\mathrm{H_2}}$ is the reactor hydrogen partial pressure defined above. At 373.15 K, the calculation used $k_H=7.68\times10^{-4}$ mol kg$^{-1}$ bar$^{-1}$, calculated from the IAPWS correlation for hydrogen in water \citep{IAPWS2004Henry}. Net hydrogen production was calculated as:
\begin{equation}
N_{\mathrm{net}}(t)=n_g(t)+n_{\mathrm{aq}}(t)-n_g(0)-n_{\mathrm{aq}}(0)+N_{\mathrm{removed}}(t),
\end{equation}
where $N_{\mathrm{removed}}(t)$ is the cumulative hydrogen removed in gas withdrawals before the sample at time $t$. No H$_2$ source term was assigned to the helium supply, and no additional gas or liquid losses were included.

At the final pressurised sample, estimated hydrogen inventories were 0.115 mmol in the headspace and 0.0144 mmol dissolved in the water. Calculated hydrogen removal during preceding withdrawals was 0.563 mmol at a sampling-loop temperature of 20$^\circ$C and 0.553 mmol at 25$^\circ$C. Subtracting the initial combined inventory of 0.0367 mmol gives estimated net production of 0.645--0.655 mmol, equivalent to 14.5--14.7 mL at STP (0$^\circ$C and 1 atm).

\section{Bulk composition and interpretation}
For bulk major-element analysis, a separate portion of PS23-011 was ground to a fine powder using a TEMA tungsten carbide swing mill. A quartered aliquot was ignited at 1100$^\circ$C to determine loss on ignition and prepared as a fusion disc. Major elements were measured using a 2010 PANalytical Axios sequential X-ray fluorescence spectrometer.

Table S3 gives the supplied major-oxide analysis of PS23-011. Total Fe reported as Fe$_2$O$_3$ is an analytical oxide convention; it does not identify Fe oxidation state, reactive Fe(II) inventory or mineral proportions. Loss on ignition does not measure organic carbon separately from mineral-bound water and other volatile components. Carbon detected in BEX spots on grains mounted on carbon adhesive likewise does not provide a bulk rock organic-carbon measurement. The bulk chemistry and microscopy provide context for H$_2$-forming reactions without uniquely identifying the reaction pathway.

\clearpage
\begin{table}[H]
\centering
\caption{Supplied bulk oxide composition of PS23-011, in wt.\%. LOI is loss on ignition. Total Fe reported as Fe$_2$O$_3$ does not determine Fe valence.}
\label{tab:bulk_oxides}
\begin{tabular}{lr}\toprule Oxide or quantity & wt.\% \\\midrule
CaO & 2.49 \\
MgO & 33.37 \\
Al$_2$O$_3$ & 4.18 \\
SiO$_2$ & 37.49 \\
P$_2$O$_5$ & 0.03 \\
SO$_3$ & 0.02 \\
Na$_2$O & 0.06 \\
K$_2$O & 0.004 \\
Fe$_2$O$_3$ & 9.353 \\
MnO & 0.14 \\
SrO & 0.001 \\
TiO$_2$ & 0.135 \\
ZnO & 0.023 \\
BaO & 0.01 \\
Cr$_2$O$_3$ & 1.01 \\
NiO & 0.198 \\
ZrO$_2$ & 0.008 \\
LOI & 11.83 \\
Reported sum & 100.35 \\
\bottomrule\end{tabular}
\par\smallskip\footnotesize The listed entries sum to 100.352 wt.\%, consistent with the rounded reported sum; values are not renormalised.
\end{table}
\clearpage
\begin{table}[H]
\centering
\caption{Companion reactor mass spectrometry data. (a) Baseline-corrected median signals divided by $10^{-12}$. (b) Reaction times, signal ratios and estimated gas mole fractions. Sample 10 was collected after depressurisation and is excluded from the pressurised time series.}
\label{tab:reactor_signals}
\footnotesize (a) Supplied signals\par\smallskip
\begin{tabular}{rrrrrrr}\toprule ID & $m/z$ 2 & 4 & 18 & 28 & 32 & 44 \\\midrule
1 & 0.060924 & 7.9246 & 1.6473 & 0.53449 & 0.094984 & 0.061993 \\
2 & 0.28143 & 8.0975 & 1.1836 & 0.2266 & 0.081497 & 0.30942 \\
3 & 0.51093 & 20.35 & 2.023 & 0.22969 & 0 & 0.6456 \\
4 & 0.23599 & 12.042 & 1.0482 & 0.097902 & 0 & 0.24068 \\
5 & 0.30542 & 15.607 & 1.398 & 0.16698 & 0 & 0.47282 \\
6 & 0.26984 & 13.956 & 1.2126 & 0.17978 & 0 & 0.4252 \\
7 & 0.40067 & 15.907 & 1.2908 & 0.21063 & 0 & 0.54869 \\
8 & 0.50659 & 17.875 & 1.5807 & 0.18989 & 0 & 0.51787 \\
9 & 2.6989 & 97.893 & 2.7206 & 0.48328 & 0 & 1.4664 \\
10 & 0.97909 & 74.002 & 2.3552 & 0.37373 & 0 & 0.76083 \\
\bottomrule\end{tabular}\par\bigskip
(b) Processed quantities\par\smallskip
\begin{tabular}{rrrrrr}\toprule ID & \shortstack{Reaction time\\(days)} & $I_2/I_4$ & $I_{44}/I_4$ & Nominal H$_2$ (\%) & Nominal CO$_2$ (\%) \\\midrule
1 & 0.00 & 0.00769 & 0.00782 & 0.2943 & 0.0945 \\
2 & 2.81 & 0.03476 & 0.03821 & 1.3334 & 0.4628 \\
3 & 5.97 & 0.02511 & 0.03172 & 0.9791 & 0.3905 \\
4 & 6.96 & 0.01960 & 0.01999 & 0.7688 & 0.2475 \\
5 & 7.85 & 0.01957 & 0.03029 & 0.7661 & 0.3744 \\
6 & 8.89 & 0.01934 & 0.03047 & 0.7571 & 0.3766 \\
7 & 12.83 & 0.02519 & 0.03449 & 0.9843 & 0.4255 \\
8 & 16.90 & 0.02834 & 0.02897 & 1.1061 & 0.3569 \\
9 & 20.75 & 0.02757 & 0.01498 & 1.0893 & 0.1868 \\
10 & 20.87 & 0.01323 & 0.01028 & 0.5257 & 0.1290 \\
\bottomrule\end{tabular}
\end{table}
\clearpage
\begin{table}[H]
\centering
\caption{Reactor sampling record. ``Failed'' denotes a sampling withdrawal without a usable mass spectrum; H$_2$ mole fractions for these events were interpolated in time between adjacent successful samples. MS ID identifies the corresponding sample in Table S4. NL denotes a usable spectrum without a pump-log entry; dashes indicate missing entries.}
\label{tab:pump_events}
\footnotesize\begin{tabular}{rrrrlrr}\toprule Event & \shortstack{Reaction time\\(days)} & \shortstack{$P$\\(bar gauge)} & Pump (mL) & Outcome & MS ID & H$_2$ used (\%) \\\midrule
1 & 0.87 & 4.1 & 411.71 & Success & 1 & 0.2943 \\
2 & 2.00 & 4.1 & 416.36 & Failed & -- & 1.0330 \\
3 & 2.00 & 4.0 & 416.49 & Failed & -- & 1.0355 \\
4 & 2.10 & 4.0 & 416.90 & Failed & -- & 1.0720 \\
5 & 2.81 & 3.8 & 421.56 & Success & 2 & 1.3334 \\
6 & 2.97 & 3.8 & 411.82 & Failed & -- & 1.3156 \\
7 & 2.98 & 3.8 & 397.72 & Failed & -- & 1.3146 \\
8 & 5.85 & 3.9 & 382.97 & Failed & -- & 0.9922 \\
9 & 6.00 & 3.8 & 382.97 & Success & 3 & 0.9791 \\
10 & 6.08 & 3.7 & 380.96 & Failed & -- & 0.9555 \\
11 & 6.96 & 4.0 & -- & NL & 4 & 0.7688 \\
12 & 7.82 & 4.0 & 352.93 & Success & 5 & 0.7661 \\
13 & 8.96 & 3.9 & 348.48 & Success & 6 & 0.7571 \\
14 & 8.97 & 4.0 & 325.27 & Failed & -- & 0.7614 \\
15 & 9.04 & 3.9 & 329.70 & Failed & -- & 0.7657 \\
16 & 11.86 & 3.8 & 321.27 & Failed & -- & 0.9282 \\
17 & 12.82 & 3.9 & 311.97 & Failed & -- & 0.9839 \\
18 & 12.83 & 4.0 & 297.45 & Success & 7 & 0.9843 \\
19 & 12.90 & 4.0 & 290.48 & Failed & -- & 0.9864 \\
20 & 12.93 & 3.9 & 309.11 & Failed & -- & 0.9873 \\
21 & 13.01 & 4.0 & 287.76 & Failed & -- & 0.9897 \\
22 & 16.90 & 3.9 & 282.98 & Success & 8 & 1.1061 \\
23 & 17.06 & 3.9 & 282.98 & Failed & -- & 1.1054 \\
24 & 20.76 & 4.0 & 268.10 & Success & 9 & 1.0893 \\
25 & 20.87 & 4.0 & -- & NL & 10 & 0.5257 \\
\bottomrule\end{tabular}
\end{table}
\clearpage
\begin{table}[H]
\centering
\caption{Experimental conditions for the micro-CT and companion reactor experiments.}
\label{tab:conditions}
\small\begin{tabular}{p{.20\linewidth}p{.34\linewidth}p{.34\linewidth}}\toprule Parameter & Imaging experiment & Companion reactor \\\midrule
Material & PS23-011 granular pack & PS23-011, nominal 200 g \\
Grain size & 0.6--1 mm sieved fraction & Crushed rock \\
Fluid & Degassed 5 wt.\% KI brine & 430 mL degassed deionised water \\
Gas boundary & No initially resolved gas; CO$_2$ preflush followed by brine flooding & Nominal 100 mL helium headspace \\
Temperature & 100$^\circ$C & 100$^\circ$C \\
Pressure & 4 bar pore pressure; 10 bar confinement & Nominal 4 bar gauge; 3.8--4.1 bar gauge at retained pressurised samples \\
Pressure control & Fluid displacement without imposed through-flow & Helium supplied by a pump to maintain pressure during gas sampling \\
Duration & Approximately 30 h & Approximately three weeks \\
Gas measurement & Micro-CT gas saturation; H$_2$ detection in pump gas effluent during depressurisation; H$_2$-indicator strip in the confining sleeve, inspected after the experiment & Baseline-corrected mass spectrometer signals \\
Matched blank & Unavailable & Unavailable \\
\bottomrule\end{tabular}
\end{table}

\bibliography{references}